\documentclass[twocolumn,prl,superscriptaddress,showpacs]{revtex4}
\usepackage{epsf,graphicx,amssymb,subfig,amsmath,color}
\begin{document}
%\draft
\title{Instabilities in magnetized spherical Couette flow}

\author{Christophe Gissinger} \affiliation{Department of Astrophysical
  Sciences, Princeton University, Princeton, NJ
  08544.}\affiliation{Center for Magnetic Self-Organization in
  Laboratory and Astrophysical Plasmas, Princeton Plasma Physics
  Laboratory, Princeton University, P.O. Box 451, Princeton, New
  Jersey 08543, USA} \author{Hantao Ji} \affiliation{Center for
  Magnetic Self-Organization in Laboratory and Astrophysical Plasmas,
  Princeton Plasma Physics Laboratory, Princeton University, P.O. Box
  451, Princeton, New Jersey 08543, USA} \author{Jeremy Goodman}
\affiliation{Department of Astrophysical Sciences, Princeton
  University, Princeton, NJ 08544.}

\def\bfnabla{\mbox{\boldmath $\nabla$}}

\begin{abstract}
We report 3D numerical simulations of the flow of an electrically
conducting fluid in a spherical shell when an magnetic field is
applied. Different spherical Couette configurations are investigated,
by varying the rotation ratio between the inner and the outer sphere,
the geometry of the imposed field, and the magnetic boundary
conditions on the inner sphere. Either a Stewartson layer or a
Shercliff layer, accompanied by a radial jet, can be generated
depending on the rotation speeds and the magnetic field strength, and
various non-axisymmetric destabilizations of the flow are observed. We
show that instabilities arising from the presence of boundaries
present striking similarities with the magnetorotational instability
(MRI). To this end, we compare our numerical results to experimental
observations of the Maryland experiment, who claimed to observe MRI in
a similar setup.

\end{abstract}
\maketitle 

\section{Introduction}

Spherical Couette flow, i.e. the flow between differentially rotating
concentric spheres, has attracted a revival of interest in recent
years. It has been shown that despite the simplicity of the problem,
the flow undergoes many bifurcations as Reynolds number increases,
depending in ratios of sphere radii and rotation speeds. When one
considers an electrically conducting fluid and imposes 
magnetic field, magnetohydrodynamic effects can significantly change
the purely hydrodynamical problem and lead to new instabilities. This
problem has evident consequences in geophysical and astrophysical
contexts (stars, planetary interiors), where setups similar to
magnetized spherical Couette flow are often encountered.

This problem has also been extensively studied in cylindrical
Taylor-Couette flow, i.e. the viscous flow between two differentially
rotating concentric cylinders. This interest in magnetized conducting
fluids confined by rotating walls has been renewed in the last decade
by investigation of the magnetorotational instability (MRI) in the
laboratory. The MRI is currently the best candidate to explain angular
momentum transport in accretion disks around stars and black holes
\cite{BalbusHawley91}. Balbus and Hawley, rediscovering an instability
first studied by Velikhov\cite{Velikhov59} and Chandrasekhar
\cite{Chandra60}, have shown that a weak magnetic field can
destabilize otherwise stable Keplerian flows. The MRI eventually
yields a magnetohydrodynamical turbulent state, enhancing the angular
momentum transport and allowing inward flow, as observed in accretion
disks.

Several experiments are currently working on this instability. The
Princeton experiment has been designed to observe MRI in a
Taylor-Couette flow of liquid Gallium, with an axial applied
magnetic field \cite{Ji06}. So far, MRI has not been identified, but
non-axisymmetric modes have been observed when a strong magnetic field
is imposed \cite{Nornberg10}. The PROMISE experiment, in Dresden, is
based on a similar set-up, except that the applied field possesses an
azimuthal component. Axisymmetric traveling waves have been obtained,
and identified as being Helical MRI, an inductionless instability
different from but connected to the standard MRI \cite{Stefani06}.

   On the other hand, spherical Couette flows have been widely
   studied, through theoretical analyses, laboratory experiments, and
   numerical simulations. For instance, without magnetic field but for
   sufficiently large Reynolds numbers, it is known that the flow can
   be hydrodynamically unstable to a rich variety of axisymmetric and
   non-axisymmetric modes. When a magnetic field is applied,
   additional magnetohydrodynamic effects are generated. Hollerbach
   \cite{Hollerbach94} first shows numerically that a free shear layer
   is created in the flow when a strong magnetic field is
   imposed. This configuration was later asymptotically analyzed by
   Kleeorin et al \cite{Kleeorin97} and Starchenko
   \cite{Starchenko98}. When the inner sphere is conducting, Dormy et
   al \cite{Dormy98} discovered that imposing a dipolar magnetic field
   yields a super-rotating jet, that is, a region of fluid with
   angular velocity larger than either boundary's. This super-rotation
   was recently observed experimentally in the DTS experiment, using a
   spherical shell filled with liquid Sodium in presence of a dipolar
   magnetic field imposed by a permanent magnet inside the inner
   sphere \cite{Brito11},\cite{Nataf08}. More recently, it was shown
   that several non-axisymmetric instabilities are generated from
   these magnetized spherical Couette flows, including destabilization
   of the meridional return flow
   \cite{Hollerbach09},\cite{Travnikov11}, or from the free shear
   layers and jets produced by the magnetic field \cite{Hollerbach01},
   \cite{Wei08}.

A few years ago, it has been claimed that MRI was obtained in a
spherical Couette flow of liquid Sodium in Maryland \cite{Sisan06}. In
this experiment, in which the outer sphere is at rest and an external
axial magnetic field is applied parallel to the rotation axis,
non-axisymmetric oscillations of both velocity and magnetic fields
have been observed, together with an increase of the torque on the
inner sphere. Although the instability appears from a hydrodynamical
state already turbulent, these oscillations have been interpreted as a
signature of the MRI.

In this article, we numerically investigate magnetized spherical
Couette flow for different configurations, including a setup similar
to the Maryland experiment. In the first section, we present the
equations and the numerical method used to study this problem. In
sections II and III, we report numerical simulations with a dipolar
magnetic field applied to the flow, and with a rotating outer
sphere. We show that magnetized spherical Couette flow yield different
non-axisymmetric instabilities with a rich variety of structures and
non-linear interactions. We compare our MHD instabilities to the MRI,
  and highlight the striking similarities between both type of
  instabilities, including the enhancement of angular momentum
  transport and the transition to MHD turbulence. Finally, our
  numerical results using an axial magnetic field are directly
  compared to the Maryland experiment.

\section{ I Equations}
We consider the flow of an electrically conducting fluid induced in a
spherical shell. The aspect ratio of the inner sphere radius $r_i$ to
the outer radius $r_o$ is set to $0.35$. $\Omega_i$ and $\Omega_o$ are
respectively the angular speed of the inner and the outer sphere. The
governing equations for this problem are the Navier-Stokes equations
coupled to the induction equation :
 \begin{equation}
 \rho{\partial {\bf u}\over \partial t} + \rho\left( {\bf u}{\bf \nabla}\right)  
{\bf u}=-{\bf \nabla}P+\rho\nu{\bf \nabla}^{2}{\bf u} +{\bf j}\times{\bf B}  \ .
\label{NS}
\end{equation}
\begin{equation}
 {\partial{\bf B}\over\partial t}={\bf \nabla}\times\left({\bf u}\times{\bf B}\right)+{1\over\mu_{0}\sigma}{\bf \nabla}^2{\bf B}  \ .
\label{ind}
 \end{equation}

where $\rho$ is the density, $\nu$ the kinematic viscosity,
$\eta=1/(\sigma\mu_0)$ is the electrical resistivity, ${\bf u}$ is the
fluid velocity and ${\bf B}$ the magnetic field. Lengths are scaled by the
shell gap $l_0=r_o-r_i$, and we use the viscous time $t_0=l_0^2/\nu$
as a typical time scale. 
The magnetic field ${\bf B}$ is scaled by
$\sqrt{\rho\mu_0\eta\Omega_i}$. The problem is thus characterized by
$3$ dimensionless numbers: the Reynolds number $Re=(\Omega_i
l_0^2)/\nu$, the magnetic Reynolds number $Rm=(\Omega_i l_0^2)/\eta$
and the Elsasser number $\Lambda=B_0/\sqrt{\Omega_i\rho\mu_0\eta}$.
It is also useful to define the magnetic Prandtl number $Pm=\nu/\eta$,
which is simply the ratio between Reynolds numbers. The magnetic and
kinetic energies reported in this article are respectively scaled by
$\rho\nu\Omega_i$ and $\rho\nu^2/l_0^2$, and the torques by $\rho\nu
l_0$.
These equations are numerically integrated using the PARODY code
\cite{Dormy98},\cite{Christensen01}. In this code, velocity and
magnetic fields are decomposed in poloidal and toroidal components and
expanded in spherical harmonics. In the radial direction, a finite
differences method is used on an irregular mesh, which decreases in
geometrical progression towards the boundaries. Time stepping is
implemented by a Crank-Nicholson scheme for the diffusive terms and an
Adams-Bashforth for the non-linear terms. Depending on the values of
our dimensionless numbers, typical numerical resolutions involve
between $150$ and $250$ radial grid points and between $64$ and $150$
spherical harmonics degrees and orders.

Magnetized spherical Couette flow has been largely studied in an
interesting series of papers \cite{Hollerbach01}, \cite{Hollerbach09},
\cite{Wei08} and some of these results are confirmed by the present
work. However, these previous studies rely on the assumption that the
magnetic Reynolds number $Rm$ can be neglected for very resistive
fluids. This assumption greatly simplifies the governing equations, in
particular the time derivative can be omitted from the induction
equation (\ref{ind}). This is one of the motivations for the present
study, since induction processes could play an important role when
describing liquid metal experiments. In particular, a finite value of
$Rm$ is necessary to observe the standard magnetorotational
instability. For instance, in the Maryland experiment (initially
designed to observe dynamo action), the electrical conductivity of the
liquid Sodium yields magnetic Reynolds numbers up to $30$, meaning
that $Rm$ can no longer be neglected in the MHD equations.

\section{ II Dipolar magnetic field and global rotation}
In this section and the next one, the inner sphere is rotating such
that $\Omega_i/\Omega_o=8$. The global rotation of the system is thus
relatively weak.  This configuration is particularly interesting in
the context of MRI experiments. Indeed, in a cylindrical geometry,
this choice of the rotation ratio would correspond to a system
slightly below the Rayleigh stability criterion, but MRI
unstable. Both spheres are taken insulating and an internal dipolar
magnetic field (held by the inner sphere) is applied to the
system. The magnetic Prandtl number is set to $1$ in this section.

\begin{figure}
\vskip -5mm
\centerline{
\includegraphics[height=55mm]{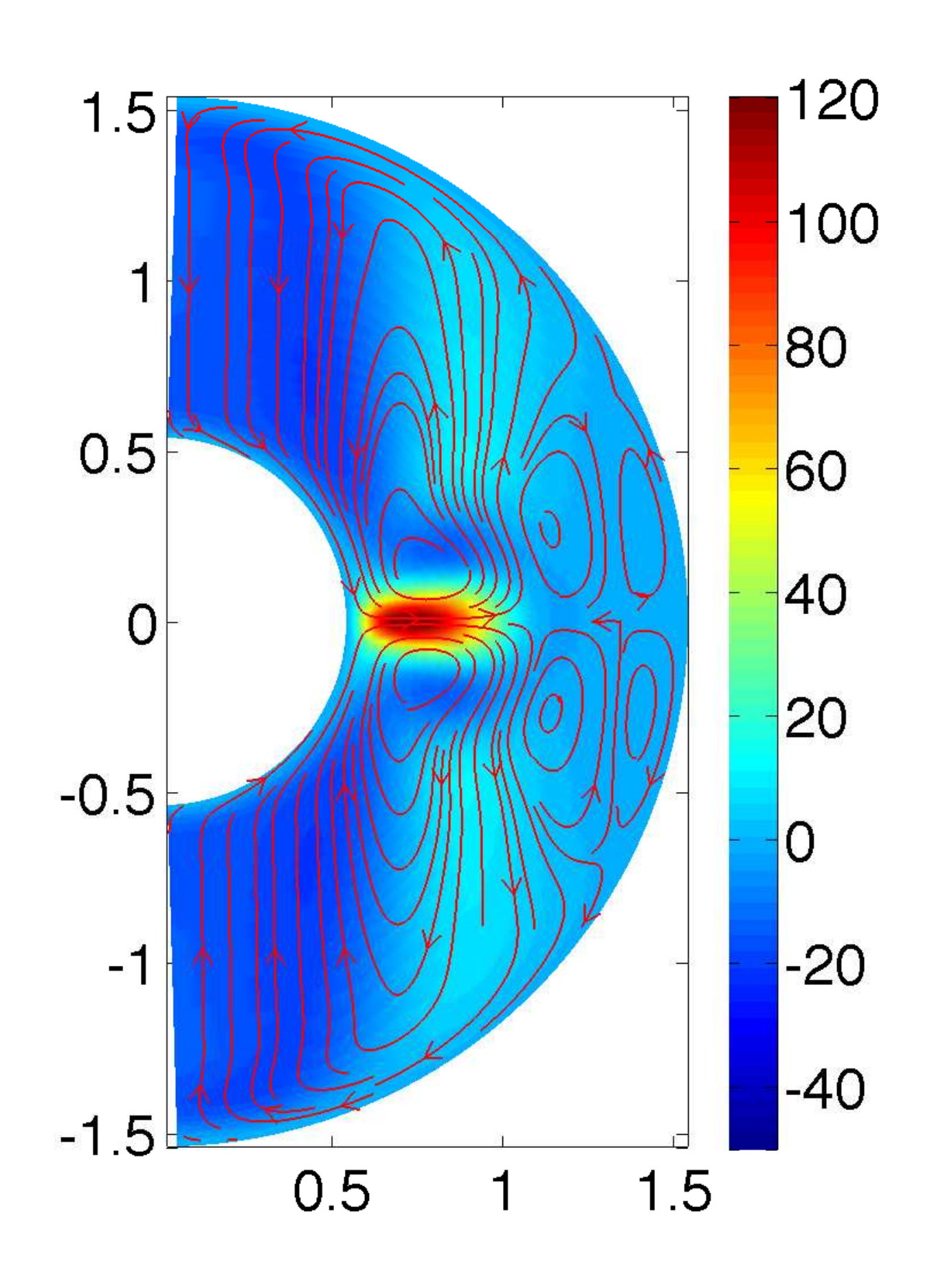} 
\includegraphics[height=55mm]{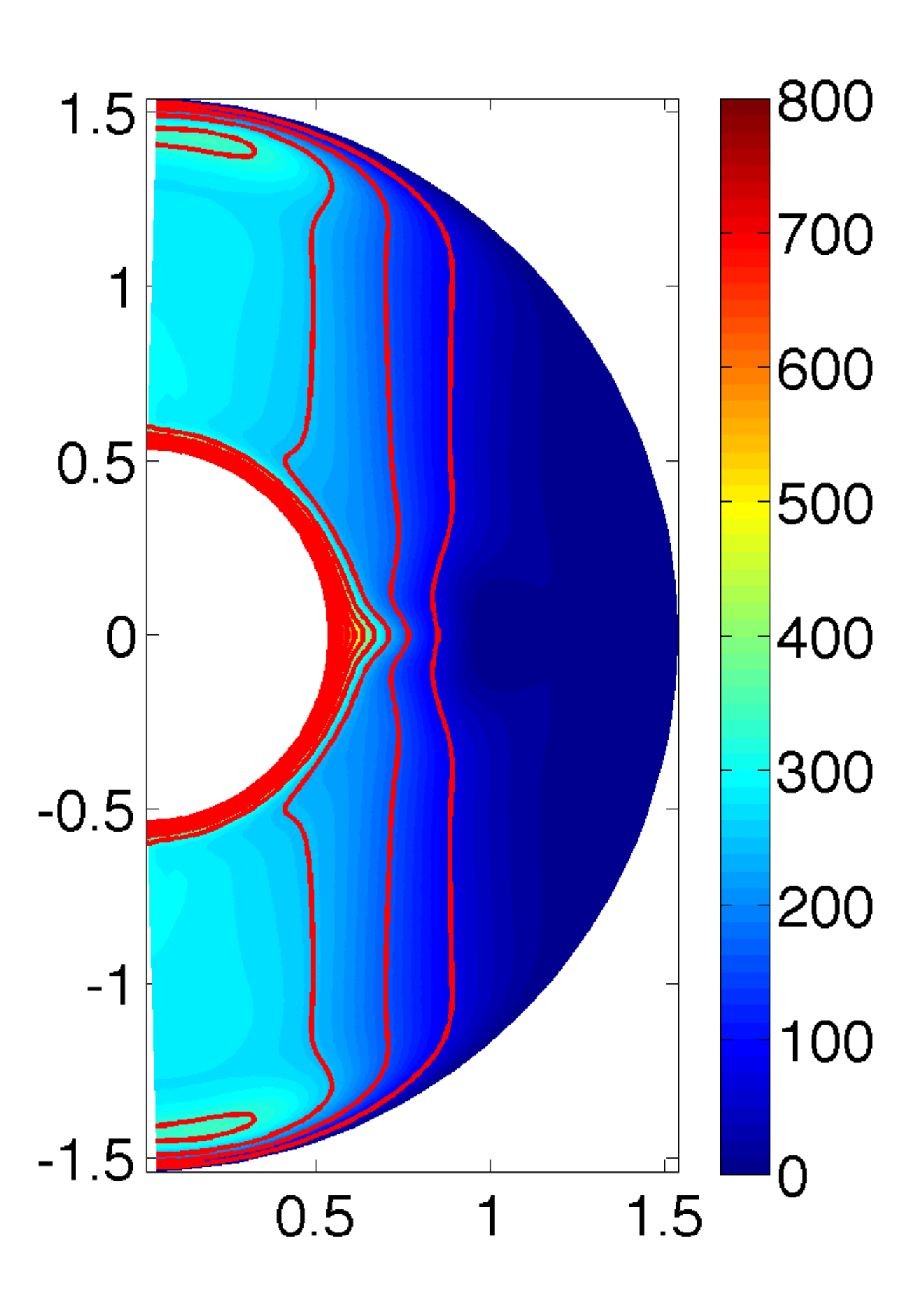} }
\caption{Purely hydrodynamical Couette flow, obtained for $Re=2000$
  and $\Omega_i/\Omega_o=8$ . Left: Radial component $u_r$ of the
  axisymmetric velocity. Note the strong equatorial jet in the
  midplane.  (lines represent streamlines of the axisymmetric poloidal
  recirculation).  Right: angular velocity in the meridional plane,
  showing a broad Stewartson layer on the tangent cylinder. Contours
  of $\Omega$ are also shown.}
\label{Couette}
\end{figure}

For small Reynolds numbers and with no magnetic field applied, the
solution is axisymmetric and corresponds to the well known spherical
Couette solution. It consists of a strong azimuthal flow associated
with a poloidal recirculation. The Taylor-Proudman theorem states that
for a sufficiently strong global rotation of the system, the velocity
tends to be uniform in the $z$ direction, due to a dominant balance
between the pressure gradient and the Coriolis force
\cite{Proudman56}. As a consequence, the flow outside the tangent
cylinder is in solid body rotation with the outer sphere. The
difference of rotation between inner and outer spheres, together with
the Taylor-Proudman constraint, yield the so-called Stewartson free
shear layer, located on the tangent cylinder
\cite{Stewartson66}. Inside the Stewartson layer, the azimuthal flow
rapidly varies from the velocity of the outer sphere (outside the
tangent cylinder) to a velocity intermediate between inner and outer
spheres (inside the tangent cylinder). The structure of the
axisymmetric component of the flow for $Re=2000$ is illustrated in
figure \ref{Couette}, and shows a Stewartson layer developing on the
tangent cylinder. Note that since these numerical simulations involve
a relatively weak global rotation, the Stewartson layer is
diffuse. Moreover, because of the differential rotation between the
inner and the outer sphere, the poloidal recirculation is
characterized by a strong equatorial jet in the midplane. A similar
jet can be observed in Taylor-Couette flow with short aspect ratio
(see for instance \cite{Kageyama04}).

\begin{figure}
\vskip -5mm
\centerline{
\includegraphics[height=72mm]{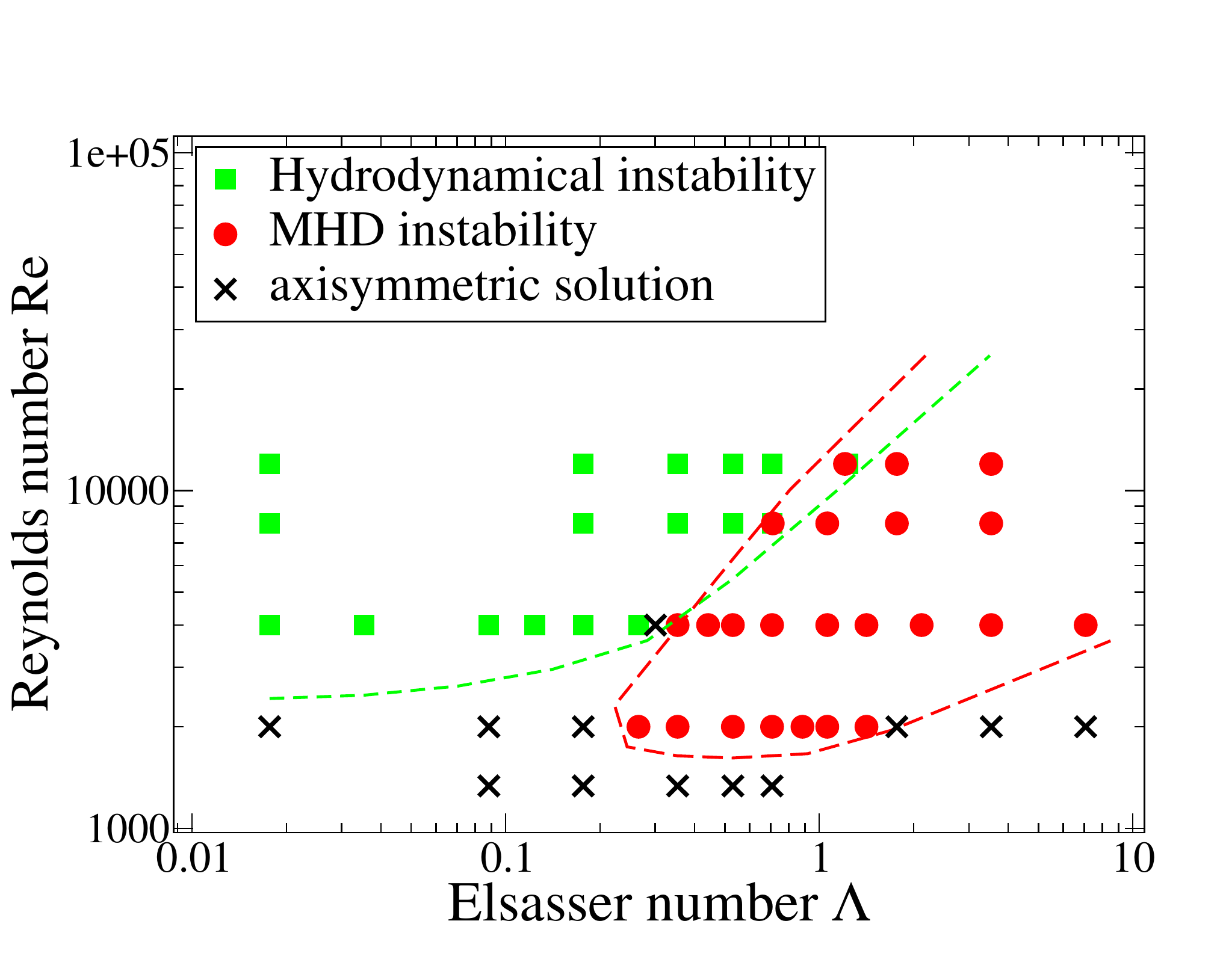} }
\caption{Parameter space for a dipolar magnetic field applied to a
  system such that $\Omega_i=8\Omega_o$, and with insulating
  spheres. Green squares indicate $m=2$ hydrodynamical instabilities
  modified (and suppressed at large $\Lambda$) by the
  applied field. Red circles are MHD instabilities of the return flow
  or the free shear layer triggered by the magnetic field, dominated
  by $m=1$ and $m=2$ azimuthal wavenumbers. Black crosses indicate
  classical spherical Couette solutions, purely axisymetric.}
\label{pri_param}
\end{figure}

For sufficiently large Reynolds number, this axisymmetric state
becomes unstable to non-axisymmetric perturbations. Figure
\ref{pri_param} shows the different states obtained in the parameter
space $(Re,\Lambda)$, and their corresponding marginal stability
curves. For $Re>2000$, the destabilization of the spherical Couette
flow yields a non-axisymmetric instability, strongly dominated by the
azimuthal wavenumber $m=2$ (green squares in figure
\ref{pri_param}).

\begin{figure}
\centerline{
\includegraphics[height=50mm]{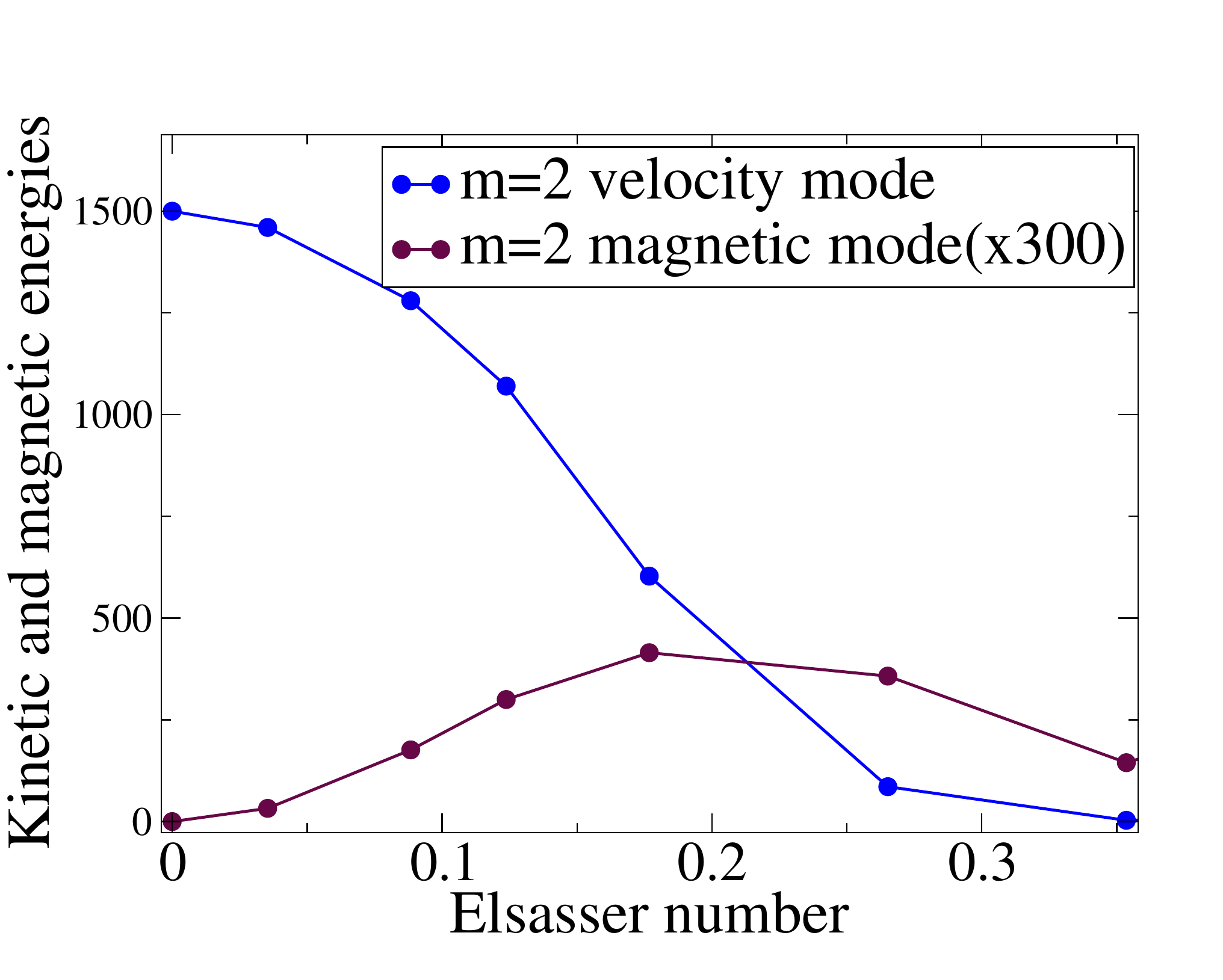}}
\centerline{
\includegraphics[height=45mm]{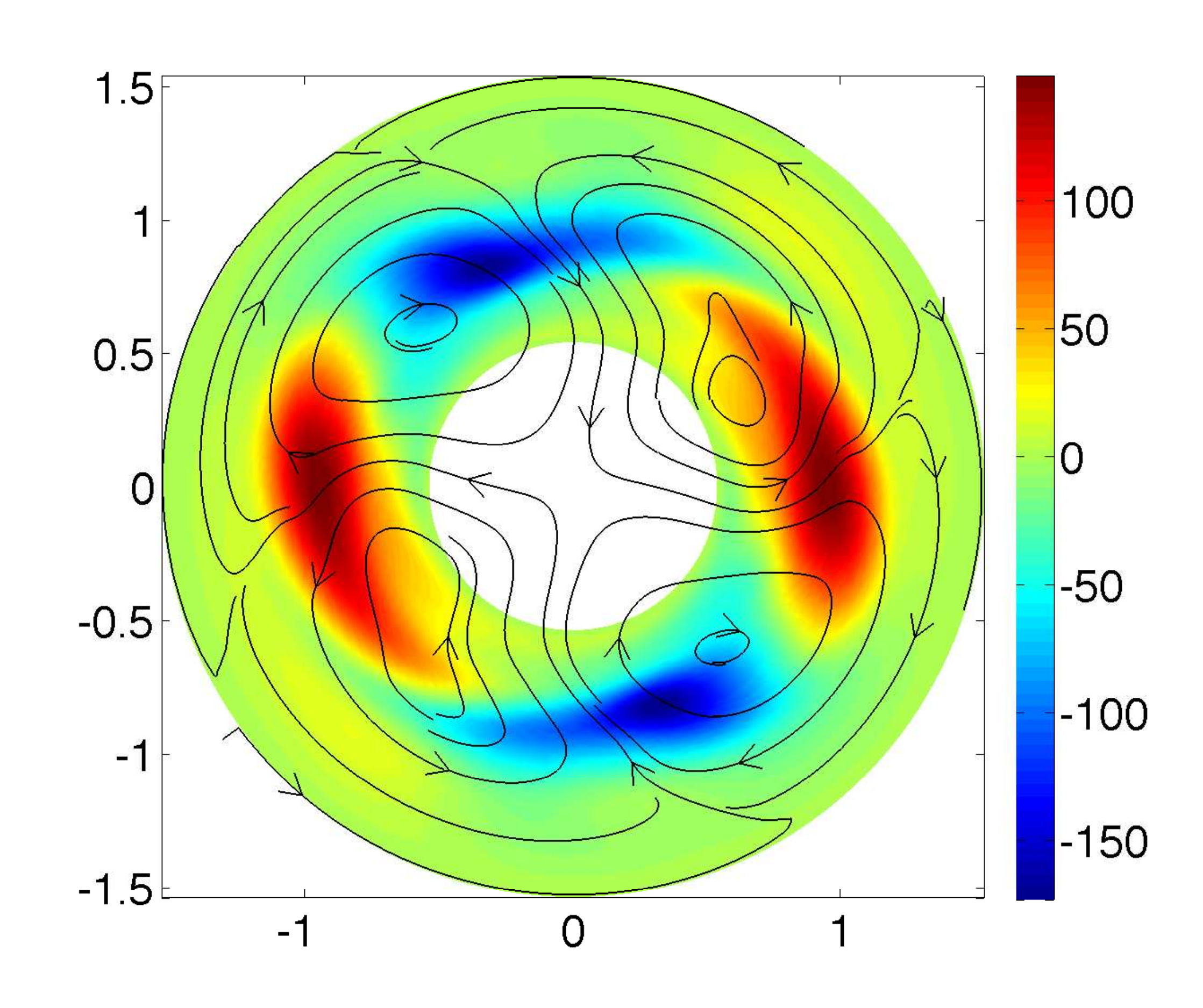} %}
\includegraphics[height=45mm]{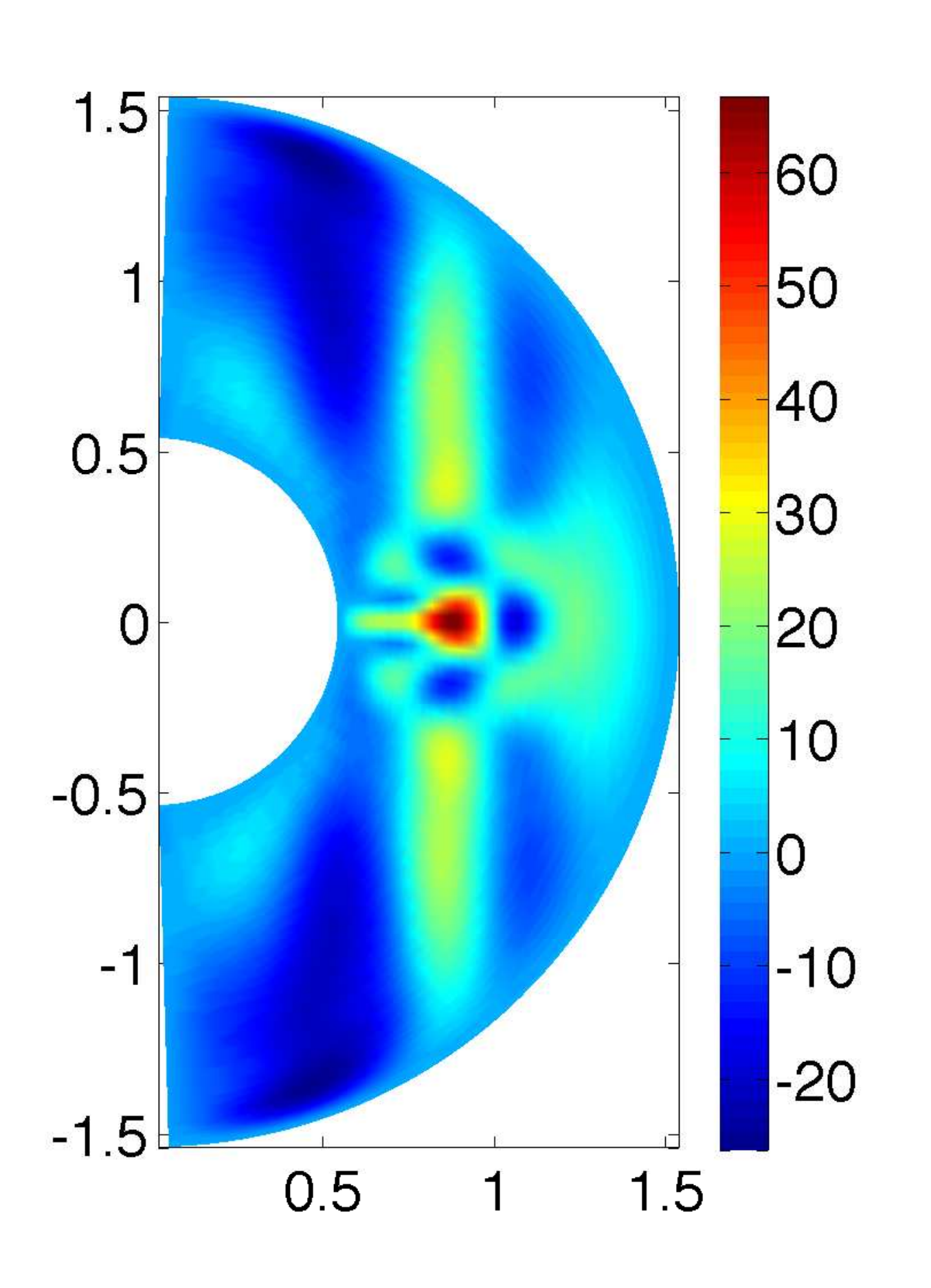} }
\caption{Top: Evolution of magnetic and velocity modes when the
  Elsasser number is increased, for $Re=4000$. The hydrodynamical
  $m=2$ non-axisymmetric instability is suppressed by the action of
  the applied dipolar magnetic field. Bottom: Non-axisymmetric $u_r$
  showing the structure of the hydrodynamical instability in the
  equatorial plane (left) and in a meridional plane at $\phi=0$
  (right), for $Re=4000$ and $\Lambda=0$. On the left are also
  indicated streamlines of the horizontal flow of the instability
  integrated in the $z$-direction.}
\label{Stew}
\end{figure}

In figure \ref{Stew} (bottom), we report the structure of this
hydrodynamical instability by showing the non-axisymmetric radial
component of the velocity field. The non-axisymmetric pattern drifts
at a constant speed in the azimuthal direction.
A cut in the equatorial plane (left) shows the $m=2$ structure of the
instability, and the meridional plane (right) illustrates its
symmetry with respect to the equatorial plane. 
Following \cite{Hollerbach09}, we denote this state by 'symmetric',
i.e. velocity and magnetic fields satisfy:
\begin{multline}
  \left(u_r, u_\theta, u_\phi, B_r, B_\theta, B_\phi\right)(r,\theta,\phi)= \\ 
\left( u_r, -u_\theta, u_\phi,-B_r,B_\theta,-B_\phi\right)(r,\pi-\theta,\phi)
\end{multline}
whereas antisymmetric modes satisfy:
\begin{multline}
  \left(u_r, u_\theta, u_\phi, B_r, B_\theta, B_\phi\right)(r,\theta,\phi)= \\ 
\left( -u_r, u_\theta, -u_\phi,B_r,-B_\theta,B_\phi\right)(r,\pi-\theta,\phi)
\end{multline}

 In such spherical Couette flows, two distinct situations are
 generally considered: configurations with a strong global rotation,
 or configurations with the outer sphere at rest. Two types of
 instability are thus observed when $Re$ is increased: In the first
 case, the Stewartson layer is destabilized and rolls up into a series
 of vortices in the $(r,\phi)$ plane, leading to an equatorially
 symmetric mode localized on the tangent cylinder. In the second case,
 if $Re$ is sufficiently large, the strong equatorial jet becomes
 unstable by adopting a wavy structure, and yields an antisymmetric
 mode localized in the equatorial plane.  In the present
 configuration, global rotation is relatively small and Reynolds
 numbers are large enough for both instabilities to be generated, so
 the interpretation is not as clear-cut. Indeed, as it can be seen in
 figure \ref{Stew}, the instability is symmetric with respect to the
 equator and consists of a series of vortices in the horizontal plane
 (i.e. the velocity perturbations are mainly horizontal), as it is
 expected from a Stewartson layer instability. Note however that a
 large part of the energy of the mode is localized in the equatorial
 plane, suggesting the the system is close to a transition to an
 equatorial jet instability.

When a magnetic field is applied to this new hydrodynamical state, the
instability can be suppressed: the flow is hydrodynamically unstable
only in a limited pocket in the parameter space (suggested by the
green dashed line in figure \ref{pri_param}). Figure \ref{Stew} (top)
shows a bifurcation diagram of the $m=2$ component of the kinetic and
magnetic energies when $\Lambda$ is increased, for $Re=4000$. It
illustrates the inhibiting role of the magnetic field: this $m=2$
non-axisymmetric mode is strongly damped by the applied field. It is
interesting to note that the observation of the magnetic energy alone
would rather suggests a destabilizing effect of the magnetic field.
For $\Lambda\sim 0.35$, the instability is completely suppressed, and
the solution is back to an axisymmetric state.

In addition to this stabilizing effect, the applied magnetic field is
also able to drive instabilities. In figure \ref{pri_param}, the red
circle indicate non-axisymmetric states, different from the
hydrodynamical instabilities described previously.

This has been interpreted as a destabilization of the poloidal return
flow by the applied field. In recent inductionless simulations,
Hollerbach \cite{Hollerbach09} has shown that in the presence of a
weak magnetic field, the poloidal return flow is destabilized to a
non-axisymmetric state. As the intensity of the magnetic field is
increased, this instability continuously connects to another type of
instability related to a free shear layer in the flow. Indeed, when a
strong magnetic field is applied to a spherical Couette flow, the
magnetic tension couples fluid elements together along the direction
of magnetic field lines. This creates a particular surface $\Sigma$ in
the flow, separating the region where magnetic field lines are
connected to both spheres from the region where magnetic field lines
are only touching one of the sphere. Depending on the region
considered, the flow will behave very differently. For instance, when
the applied field is dipolar, like in this section, $\Sigma$ is
defined by magnetic field lines just touching the outer sphere at the
equator. In this case, the fluid inside $\Sigma$, coupled only to the
inner sphere, co-rotates with it, whereas fluid outside $\Sigma$
rotates at an intermediate velocity (except near the sphere
boundaries). The jump of velocity on the surface $\Sigma$ therefore
results in a new free layer, the so-called Shercliff layer
\cite{Shercliff53}. This effect was first described in spherical
geometry by Starchenko \cite{Starchenko98}, who found that the
thickness of this layer scales like $(\Lambda\sqrt{Re})^{-1/2}$. Like
the Stewartson layer, the Shercliff layer becomes unstable to
non-axisymmetric perturbations when the shear is strong
enough. However, in the case of a dipolar applied field, the
significance of the field lines $\Sigma$ can be completely eliminated
if the Reynolds number is large enough \cite{Hollerbach07}, and only
instabilities of the return flow will remain.
Despite the differences in the geometry of the applied field and in the
 parameter regime, it is surprising to note that our figure
 \ref{pri_param} is very similar to Fig.$2$ of \cite{Hollerbach09}.

\begin{figure}
\includegraphics[height=60mm]{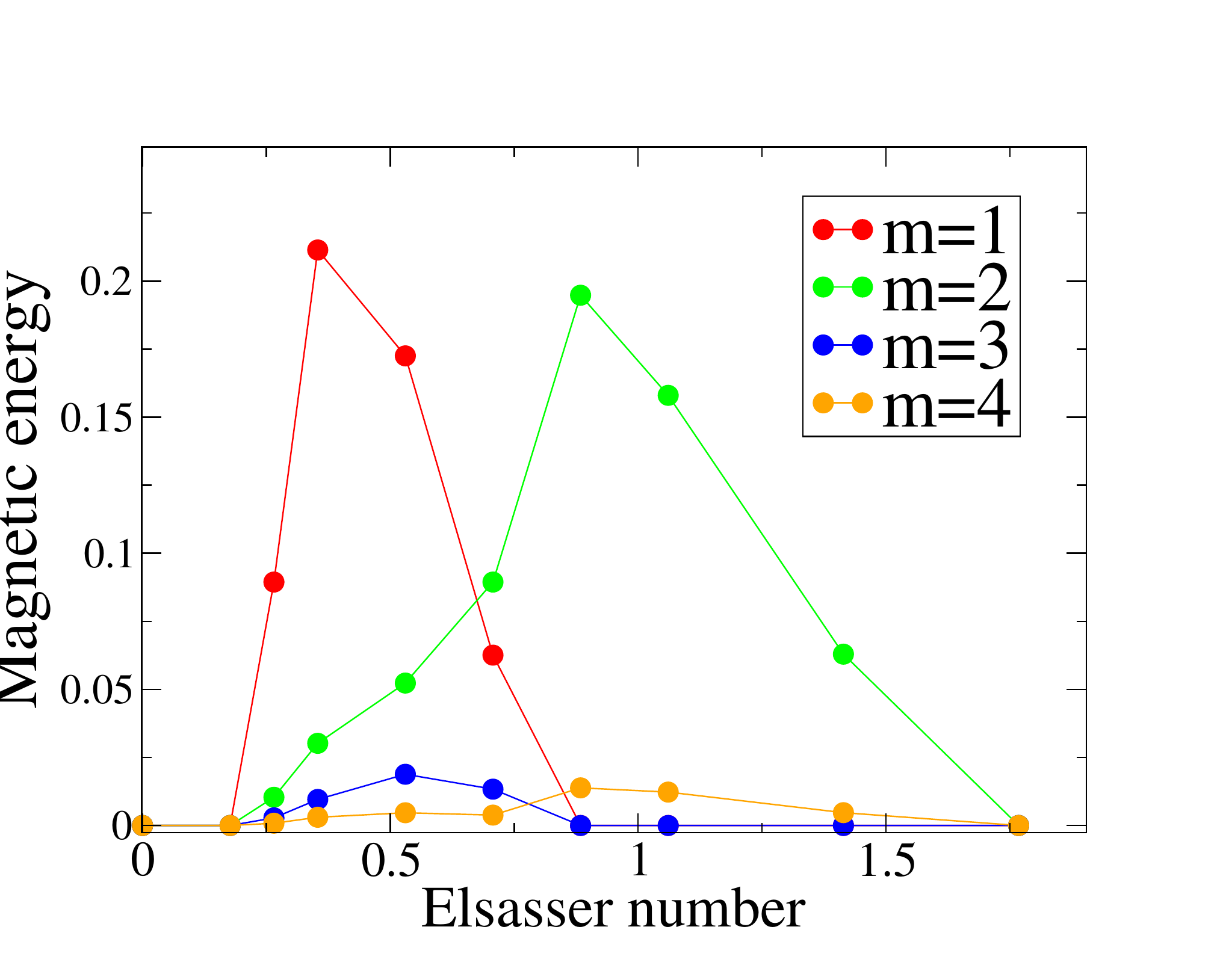}
\caption{Bifurcation diagram of the magnetic energy when the Elsasser
  number $\Lambda$ is increased for a fixed Reynolds number
  $Re=2000$. Note the non-axisymmetric destabilization of the
  meridional return flow by the applied field, dominated by $m=1$ and
  $m=2$ non-axisymmetric modes. Strong applied fields suppress these
  MHD instabilities.}
\label{ScherB}
\end{figure}

Figure \ref{ScherB} shows the evolution of the non-axisymmetric
components of the magnetic field when the Elsasser number is
increased, for $Re=2000$. For this small Reynolds number, both
Stewartson layer and equatorial jet are hydrodynamically stable to
non-axisymmetric perturbations for $\Lambda=0$. As $\Lambda$ is
increased from this axisymmetric state, the return flow becomes
  unstable, and the system undergoes a bifurcation to an $m=1$
  rotating mode at $\Lambda_c=0.2$ (corresponding to red dots in
  figure \ref{pri_param}).

\begin{figure}
\centerline{
\includegraphics[height=45mm]{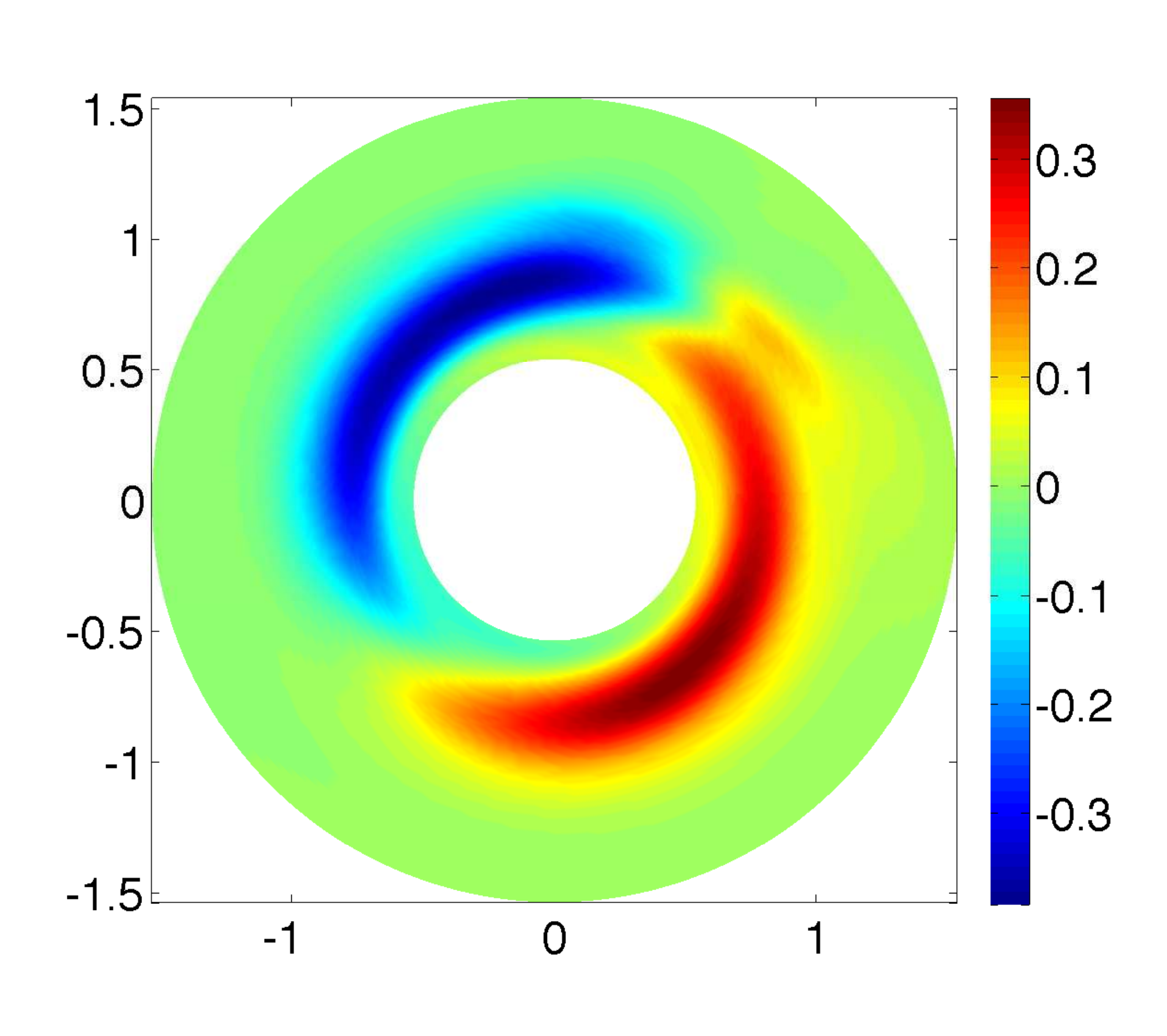} 
\includegraphics[height=45mm]{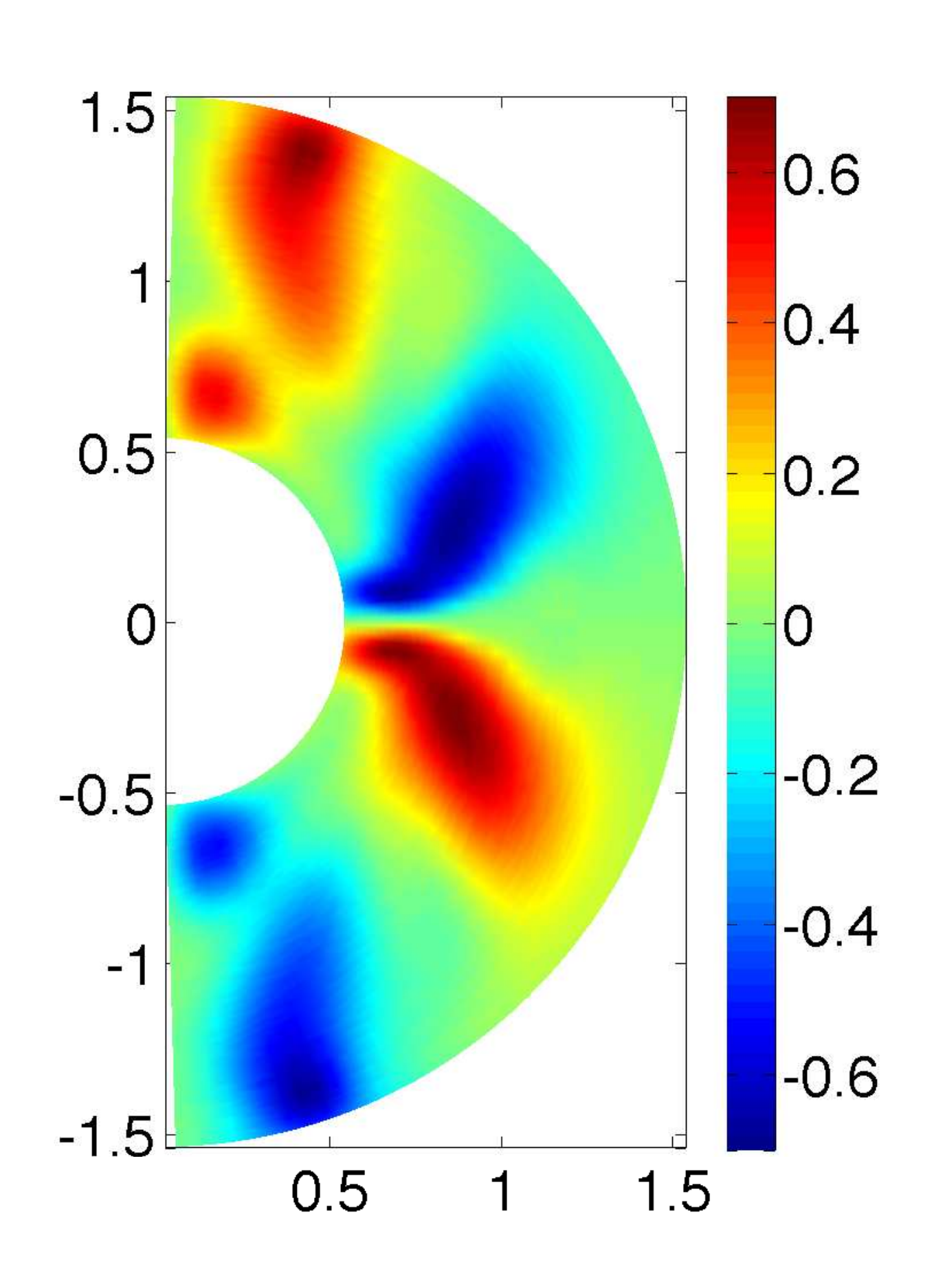} }
\centerline{
\includegraphics[height=45mm]{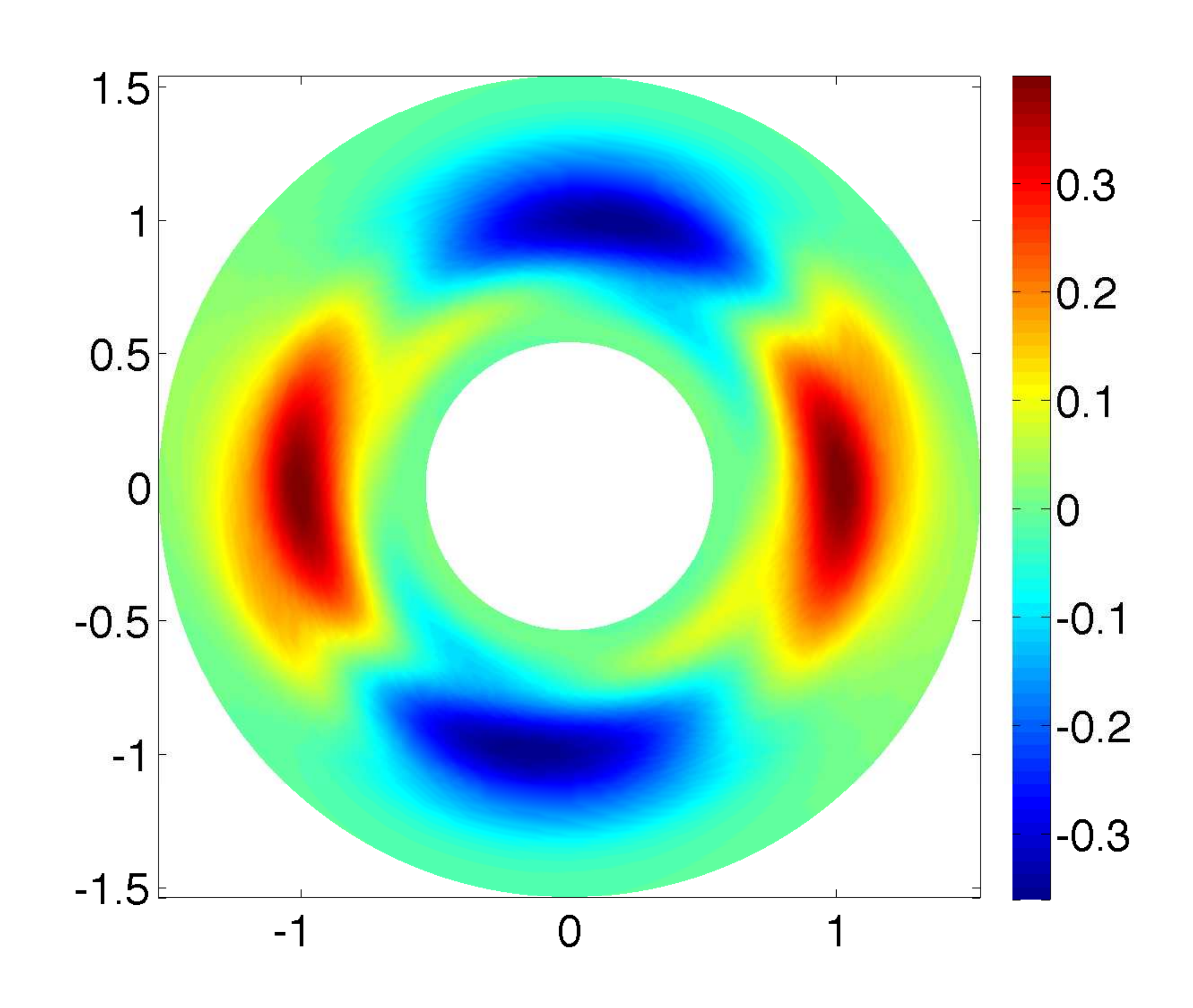} 
\includegraphics[height=45mm]{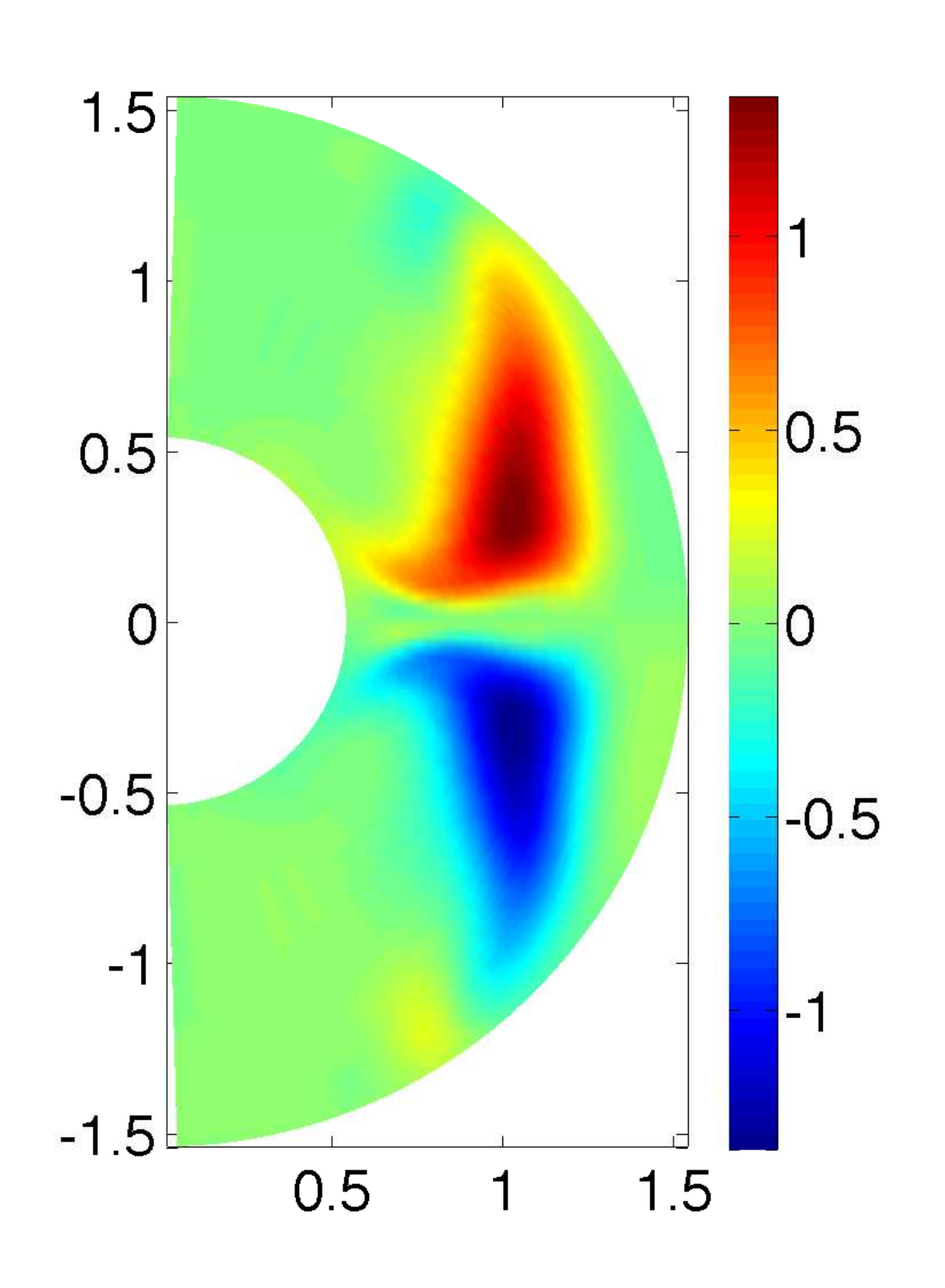} }
\caption{Structure of the MHD instability for
  $Re=2000$. The figure shows the radial non-axisymmetric component of
  the magnetic field just above the equatorial plane
  (left) or in a given meridional plane at $\phi=0$(right). The two
  top figures show the $m=1$ mode obtained for $\Lambda=0.35$ and the
  two bottom figures show the field obtained for $\Lambda=0.9$,
  dominated by the $m=2$ azimuthal wavenumber.}
\label{Struc_m1}
\end{figure}

When $\Lambda$ is increased further, one observes a transfer between
this $m=1$ mode and an $m=2$ structure. Figure \ref{Struc_m1} shows
that the modes are still symmetric with respect to the
midplane. 
The structure of the instability is similar to the one obtained in
\cite{Hollerbach09} and the localization of the energy of the mode
indeed suggests that the instability is related to the meridional
return flow. Note however that close to the threshold, our large $Rm$
calculations lead to the generation of an $m=1$ mode, whereas
inductionless simulations of \cite{Hollerbach09} predicted an $m=2$
mode.

 When $m=1$ and $m=2$ azimuthal modes are both excited, a phase
 locking is achieved between these two modes, and nonlinear
 interactions can become important. Finally, if the Elsasser number is
 increased up to $1.8$, the magnetic tension of the applied magnetic
 field suppresses any instability, and the system comes back to an
 axisymmetric state.

\section{III Comparison with MRI}
At this point, we would like to briefly compare these instabilities,
mainly due to the presence of boundaries, to what
would be expected in a cylindrical geometry. Chandrasekhar
\cite{Chandra60} and Velikhov \cite{Velikhov59} first studied the
effect of an axial applied magnetic field on a Couette flow confined
between two infinite cylinders. They considered rotations of the
cylinders such that the flow is stable under the centrifugal
instability. According to the Rayleigh criterion, this means that the
rotation profile follows $\Omega\sim r^\gamma$, with
$\gamma>-2$. However, this flow can be destabilized by applying an
magnetic field to the system if $\gamma<0$. This powerful
magnetorotational instability (MRI), later rediscovered by Balbus and
Hawley in the framework of accretion disks, yields a radial outflow of
angular momentum and can lead to an MHD turbulent state. The magnetic
field can also have a stabilizing effect: for unstable flow satisfying
$\gamma<-2$, the magnetic field is able to suppress the centrifugal
instability \cite{Chandra61}. Finally, the MRI is also suppressed if
the applied magnetic field is too strong.

Despite the fact that MRI is generally considered as a local
instability (without regard to the boundaries of the system) , it is
interesting to note that most of the above features are encountered in
the instabilities of our finite geometry system. First, figure
\ref{pri_param} shows that the MHD instability is
excited only in a delimited pocket in the parameter space: it requires
a finite value of the magnetic field to be unstable, but is stabilized
if the applied magnetic field is too strong. Second, the critical
value of the Elsasser number for restabilization rapidly increases
with $Rm$, similarly to the magnetorotational instability. In
addition, one can find similarities between the suppression of the
hydrodynamical instability by the magnetic field and
the suppression of the centrifugal instability in Taylor-Couette flow.

\begin{figure}
\centerline{
\includegraphics[height=35mm]{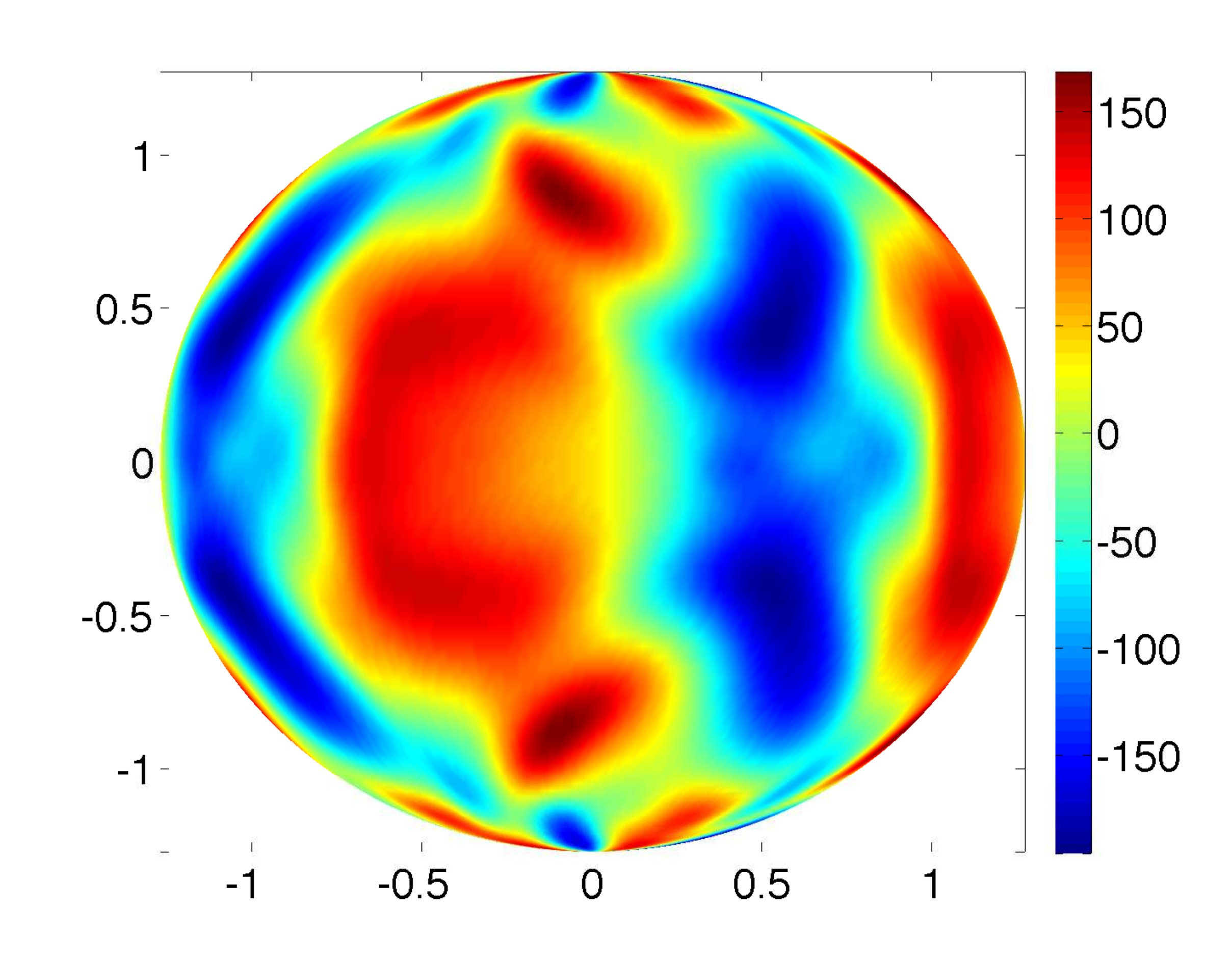} 
\includegraphics[height=35mm]{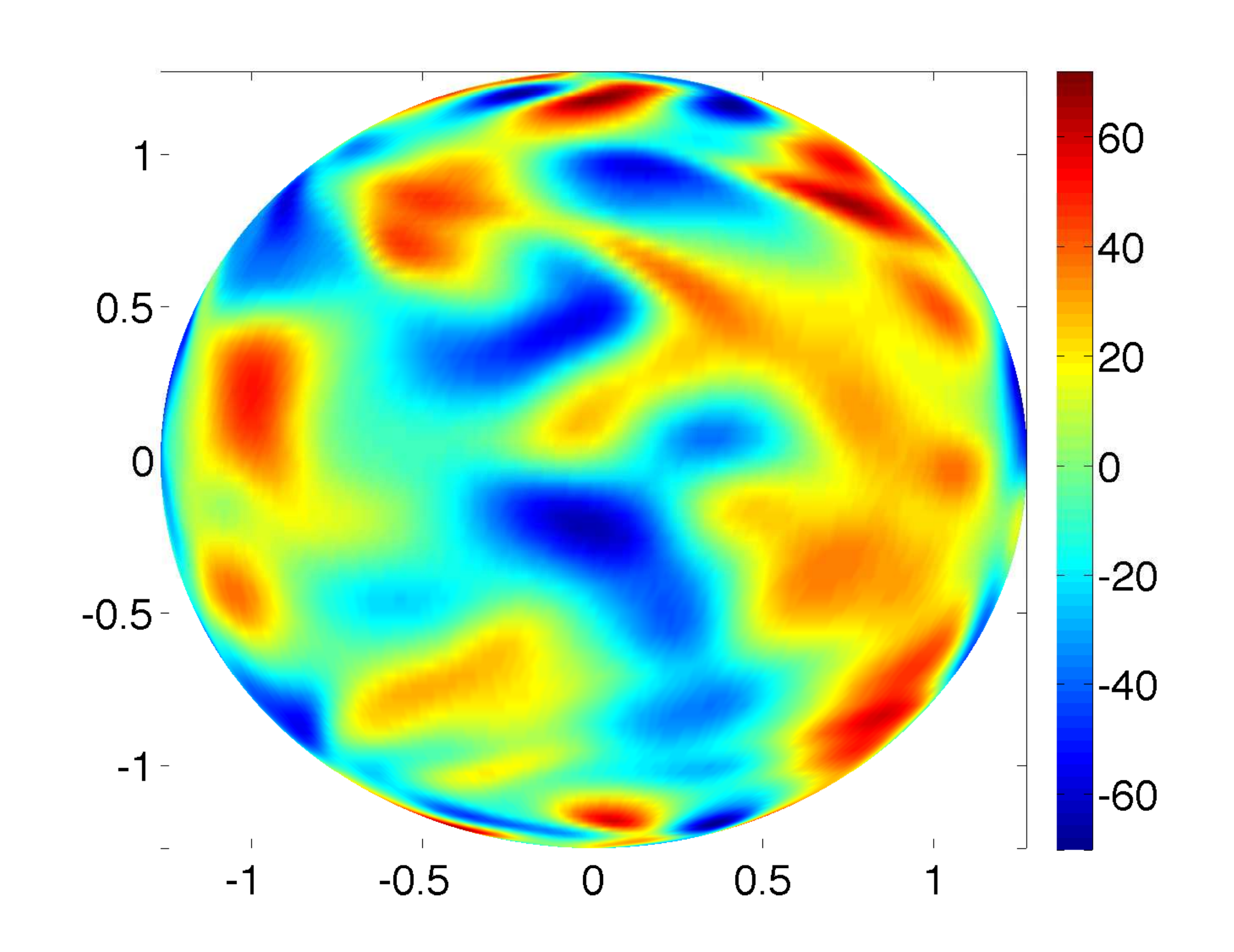} }
\centerline{
\includegraphics[height=50mm]{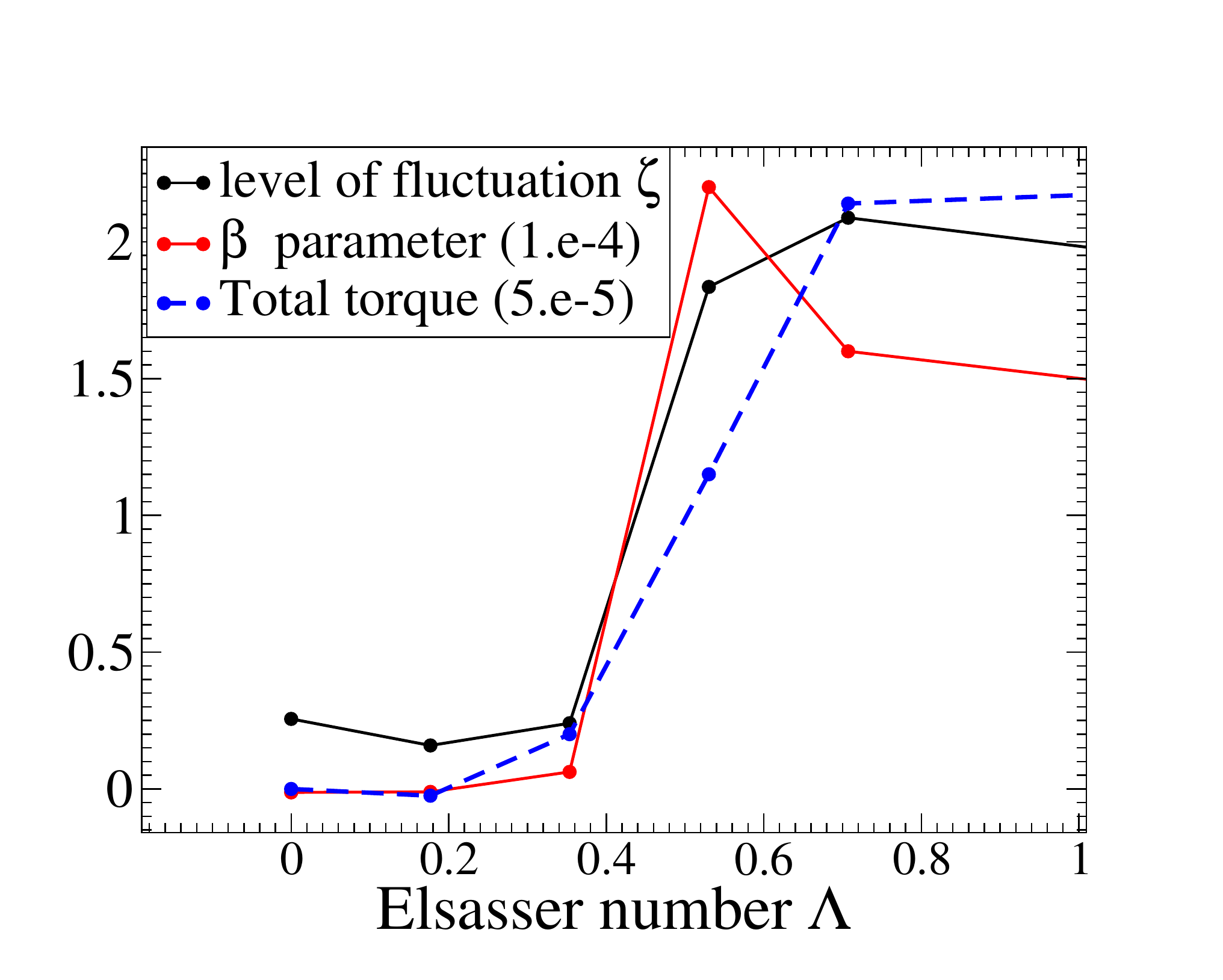} }
\caption{Effect of the imposed magnetic field on the turbulence level,
  for $Re=8000$ and $\Omega_i=8\Omega_o$. Top: non-axisymmetric
  $u_\phi$ at the spherical surface of radius $r=1.2$ showing that for
  $\Lambda=0.$ (left), the flow is dominated by an $m=2$ mode due to
  an hydrodynamic instability, and is weakly fluctuating. Right: For
  $\Lambda=1$, when boundary-driven instabilities are generated by the
  applied field, the flow becomes more complicated and
  chaotic. Bottom: evolution of the turbulence intensity when the
  magnetic field is increased. MHD instabilities due to boundaries
  yield turbulent fluctuations and angular momentum transport,
  similarly to the MRI.}
\label{turb}
\end{figure}

From an astrophysical point of view, an important characteristic of
the MRI is its ability to ensure an efficient outward transport of the
angular momentum and to yields accretion, and eventually
turbulence. Since there are different ways to measure the level of
fluctuations in our simulations, we compare three different
quantities. For instance, we computed the quantity
$\zeta={\sqrt{<(u_\phi-<u_\phi>)^2>}\over\sqrt{<u_\phi>^2}}$,
measuring the level of fluctuation of $u_\phi$, where $<>$ denotes
time averaging. The velocity is probed in the equatorial plane at
$r=0.7$ and $\phi=0$. Figure \ref{turb} shows the evolution of $\zeta$
when the Elsasser number is increased, for a fixed value of the
Reynolds number $Re=1000$. In figure \ref{turb}, we also show the
evolution of the parameter
$\beta={<(u_\phi-<u_\phi>)(u_r-<u_r>)>\over<u_\phi>^2}$ related to the
Reynolds stress, and the evolution of the excess torque $G$ applied on
the inner sphere, which quantify the amount of angular momentum
transported. For this value of $Re=8000$, the purely hydrodynamical
state consists of a basic spherical Couette flow associated with an
$m=2$ component, and the total kinetic energy is steady. As $\Lambda$
is increased, the flow becomes unstable to several
azimuthal wavenumbers, and non-linear interactions rapidly lead to a
chaotic evolution of the flow.  As can be seen on figure
\ref{turb}, this corresponds to a growth of the three quantities
$\zeta$, $\beta$ and $G$, evidencing an increase of the level of
turbulence and of the amount of angular momentum transported
outward. This transition is illustrated by the two snapshots of figure
\ref{turb}: for $\Lambda=0$ (left), the radial flow taken at
$r=(r_o+r_i)/2$ is dominated by a smooth $m=2$ perturbation. For
$\Lambda=1$ however (right), far from the MHD
instability onset, the velocity field appears more fluctuating and
spatially disorganized. It is remarkable that the three different
quantities used to measure this transition show a good agreement. This
is particularly interesting for comparison between different
experiments, where only either torque or velocity measurements are
often available.

Here again, we find that this instability plays a role similar to the
MRI, by destabilizing an otherwise stable flow and leading to a strong
MHD turbulent state. These striking similarities between our
instability and the magnetorotational instability have important
consequences for experimental studies of the MRI. In particular, it
points out the difficulties to make a clear distinction between both
types of instability from experimental diagnostics. In the next
section, we therefore compare our numerical simulations to
experimental results obtained in the Maryland experiment, which have
been interpreted as non-axisymmetric MRI.

\section{ IV Comparison with experiments: axial field and outer sphere at rest}
In this section, we now keep the outer sphere at rest, and the applied
magnetic field is purely axial, aligned with the axis of
rotation. While the outer sphere is always taken insulating, two types
of boundary conditions have been used for the inner sphere: insulating
or conducting (with the same electrical conductivity as the fluid).
This configuration is similar to the one used in the Maryland
experiment. This experiment consists of Sodium flowing between an
inner sphere of radius $a=0.050$ m and an outer sphere of radius
$b=0.15$ m. The inner sphere is made of high conductivity copper, and
an external axial magnetic field is applied coaxially using a pair of
electromagnets. In \cite{Sisan06}, the authors report that
axisymmetric and non-axisymmetric modes are spontaneously excited for
sufficiently large $Rm$ when a magnetic field is applied. Most of the
parameter space is dominated by an $m=1$ precessing pattern, and the
appearance of these instabilities is correlated with a strong increase
of the torque applied on the inner sphere. Some of the
non-axisymmetric modes are suppressed for large magnetic field. These
results have been interpreted by the authors to be a signature of the
magnetorotational instability. Note that since the outer sphere is at
rest, the background flow is already very turbulent without applied
field, in contrast with the initial stable laminar state generally
considered in studies of MRI.

\begin{figure}
\centerline{
\includegraphics[height=50mm]{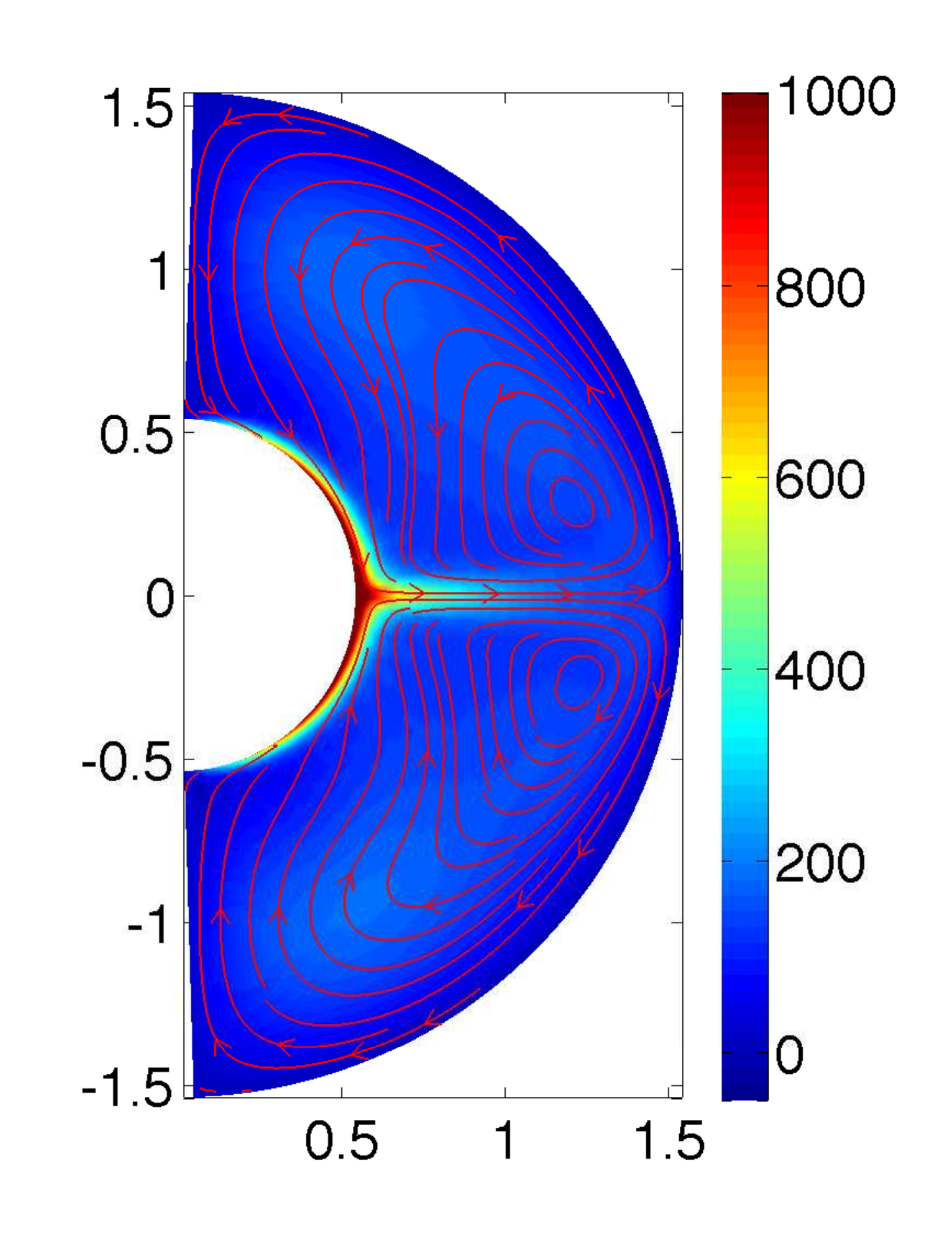}
\includegraphics[height=50mm]{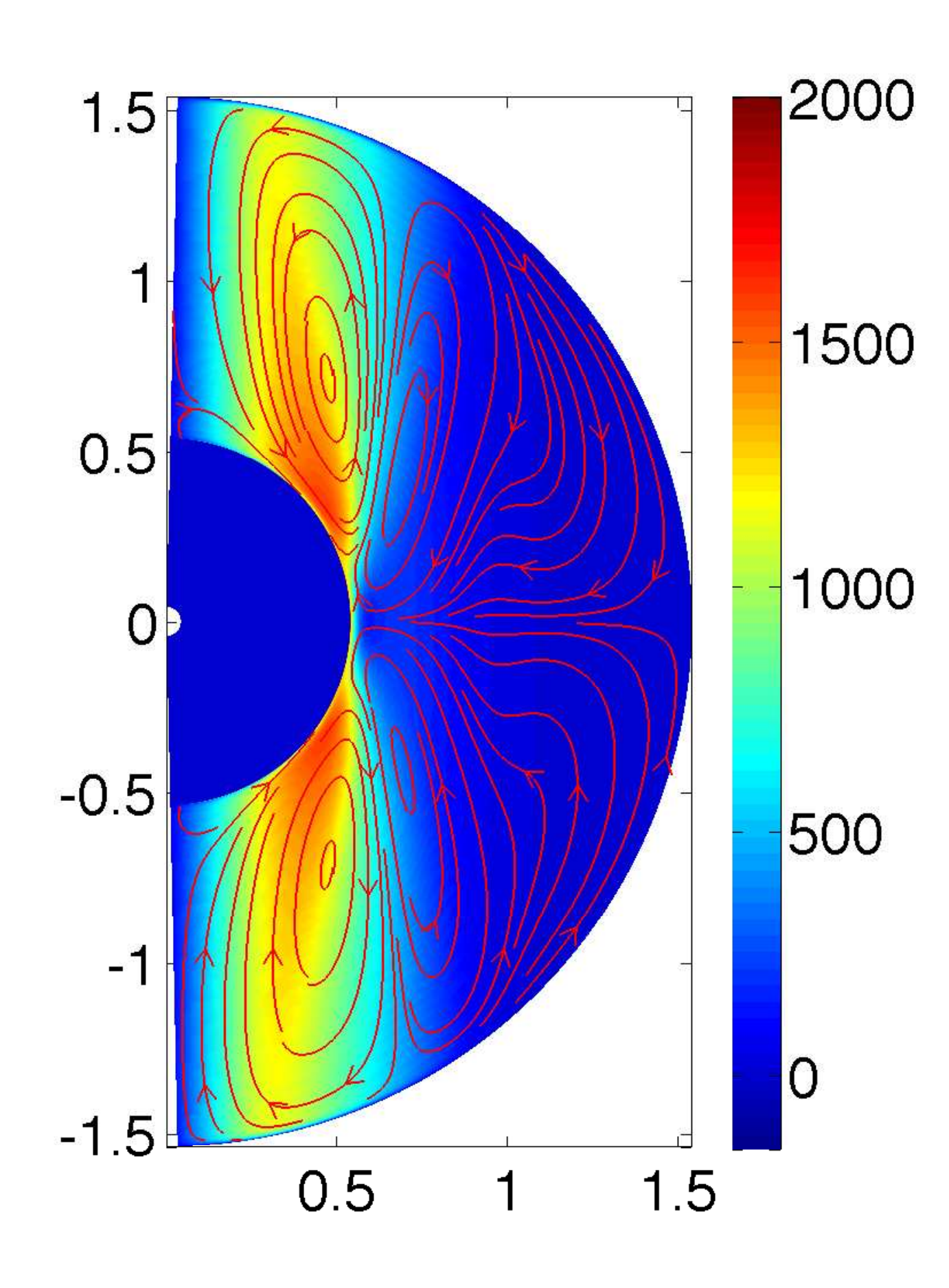}}
\centerline{
\includegraphics[width=60mm]{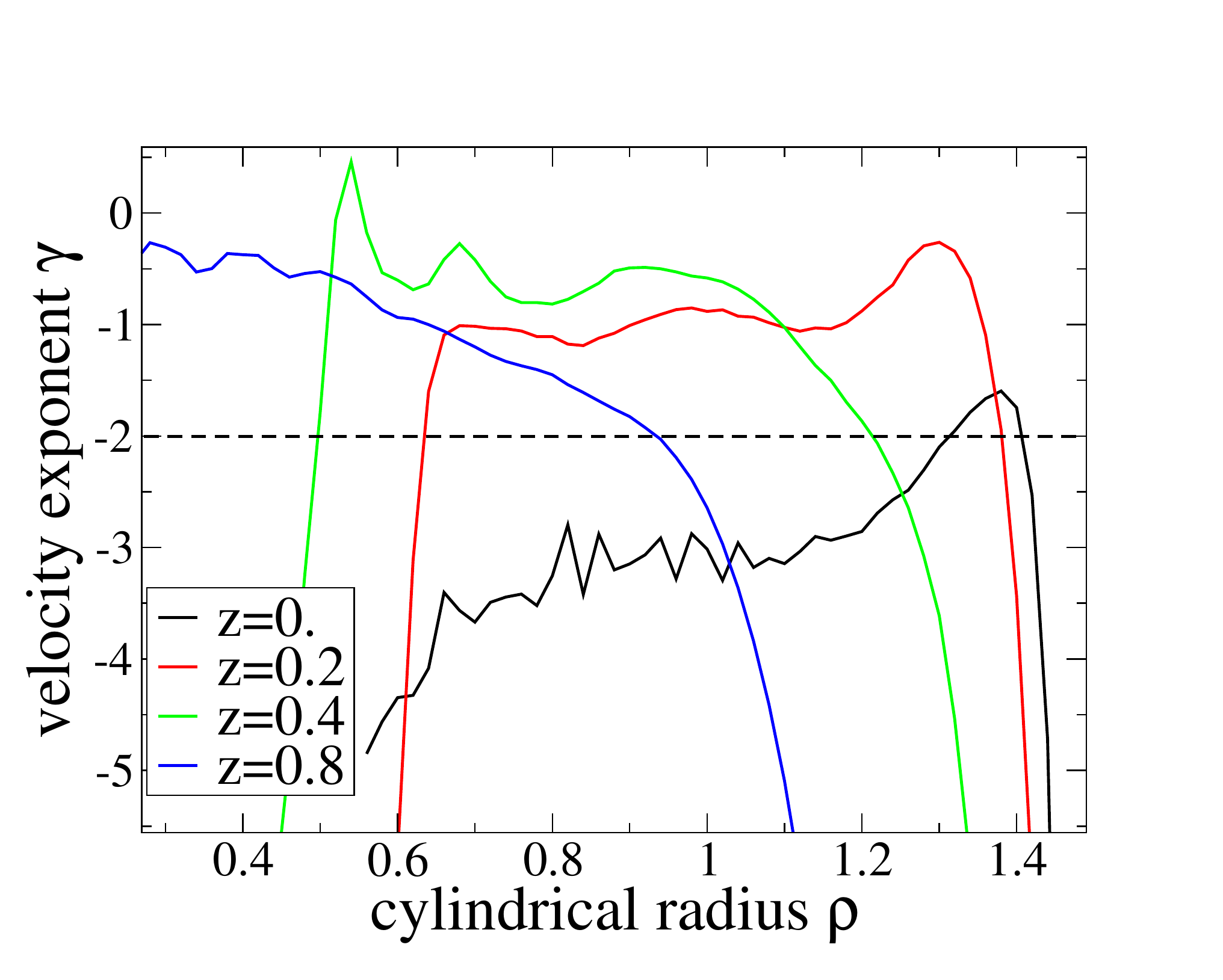}}
\caption{Top: Structure of the spherical Couette flow when the outer
  sphere is now at rest, for $\Lambda=0$ (left) and $\Lambda=2$
  (right), with $Re=5000$. Colors indicate azimuthal flow $u_\phi$ and
  streamlines show the meridional recirculation. Note the absence of
  Stewartson layer in the purely hydrodynamical case (no global
  rotation) and the generation of the Shercliff layer in the
  magnetized one. Bottom: Velocity exponent $\gamma$ taken at
  different altitude $z$, for $\Lambda=0$. Note the Rayleigh unstable
  profiles ($\gamma<-2$) close to the midplane.}
\label{sis0}
\end{figure}

As can be seen in figure \ref{sis0}, the flow without magnetic field
is very different from the one obtained in the previous sections, when
the outer sphere was rotating: a strong equatorial jet is produced in
the midplane, and no Stewartson layer is generated. It is useful to
introduce the velocity exponent $\gamma=\partial \log\Omega(r)/\partial
\log r$, where $\Omega(r)$ is the rotation rate at a cylindrical
radius $r$. The Rayleigh criterion for stability, $\gamma>-2$, is thus
expected to be violated when the outer sphere is at rest. However, in
\cite{Sisan06}, a very weak level of turbulent fluctuations has been
reported, together with a velocity exponent around $-1.5$,
surprisingly close to a stable keplerian flow. In figure \ref{sis0},
we show the radial profile of $\gamma(r)$, for different altitudes $z$
in a purely hydrodynamical simulation. Except near the boundaries
(where the flow is strongly Rayleigh unstable), stable profiles with
$\gamma>-2$ can be obtained, depending on the altitude $z$ (for
instance, $\gamma \sim -1$ at $z=0.2$ ). However, as the measurements
are done closer to the midplane, the velocity exponent significantly
decreases, and unstable profiles are obtained, with $\gamma$ much
smaller than $-2$. This is also the case if $z$ is too large. This
variability of the angular profile underlines the difficulties of
studying the MRI in a spherical Couette flow, particularly in the
absence of global rotation uniformizing the flow in the axial
direction.

\begin{figure}
\centerline{
\includegraphics[height=60mm]{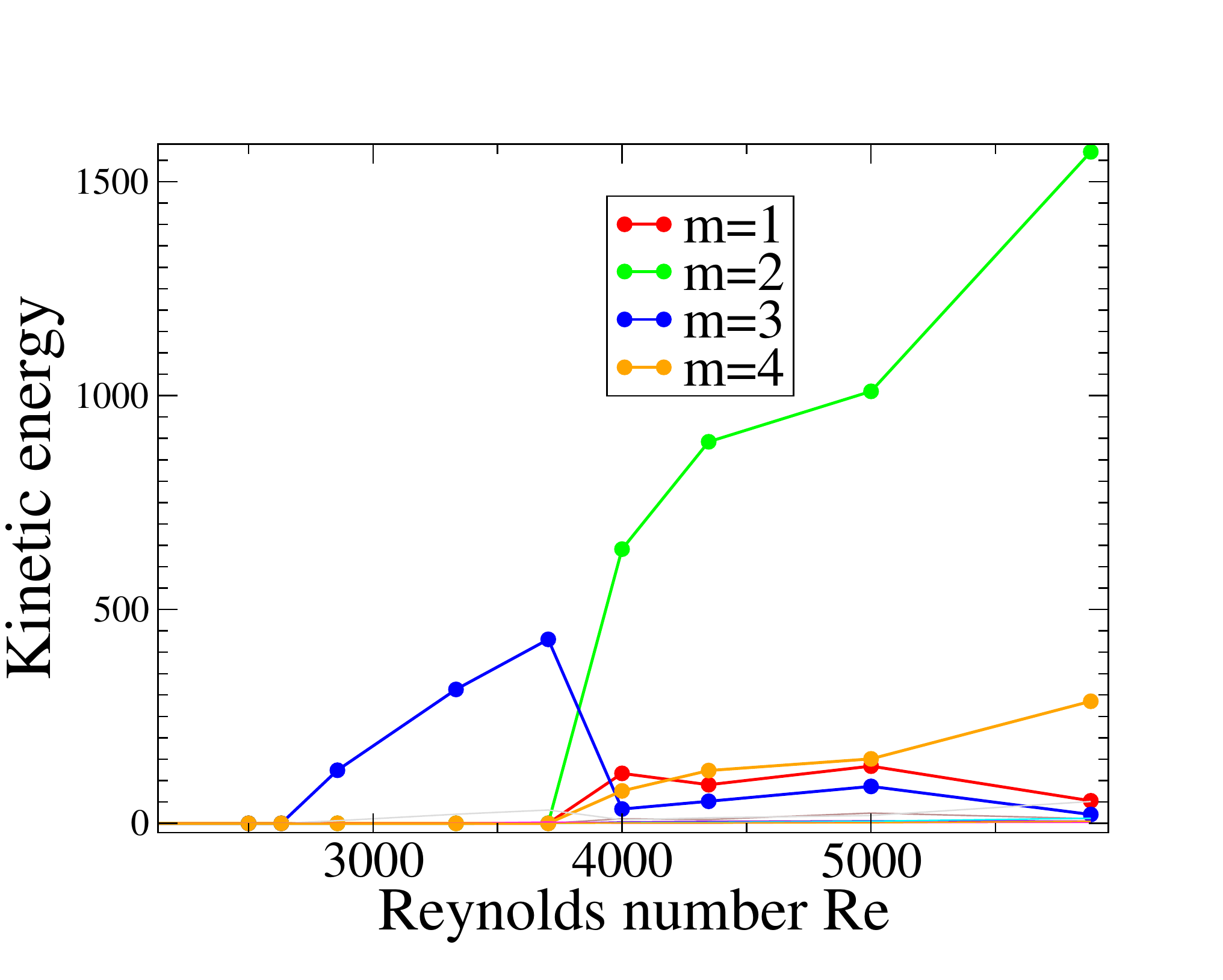}}
\caption{Bifurcation diagram of the total kinetic energy when $Re$ is
  increased. The outer sphere is at rest and Elsasser number is set to
  zero. As $Re$ is increased, the equatorial jet adopts a wavy
  structure corresponding to non-axisymmetric modes, first $m=3$ then
  $m=2$. For larger $Re$, it is expected that the flow tends to a
  fully turbulent state.}
\label{bif_hydro_sis2}
\end{figure}

 As $Re$ is increased, the flow undergoes bifurcations to
 non-axisymmetric modes. Indeed, it is well known that for
 sufficiently large Reynolds number, the equatorial jet becomes
 unstable to non-axisymmetric perturbations, and gives rise to a
 Kelvin-Helmoltz instability. The critical wavenumber for this
 instability depends on the aspect ratio and on the Reynolds
 number. Figure \ref{bif_hydro_sis2} shows a bifurcation diagram of
 the kinetic energy as $Re$ is increased. The flow bifurcates to a
 non-axisymmetric state at $Re=2700$, where the equatorial jet is
 destabilized to an $m=3$ structure. For higher $Re$, there is a
 transition to an $m=2$ mode. It is numerically expensive to conduct
 simulations at higher Reynolds numbers, but it is expected that
 successive bifurcations involving higher azimuthal wavenumbers
 eventually yield a turbulent state. It has been recently shown that
 these non-axisymmetric instabilities can trigger dynamo action, but
 only when $Pm>1$ \cite{Guervilly10}, which will not be considered
 here. An important feature is that these non-axisymmetric modes are
 antisymmetric, in contrast with the symmetric hydrodynamical modes
 obtained when the outer sphere is rotating.

Let us now study the magnetized regime. In this section, the magnetic
Prandtl number is set to $Pm=0.01$, which allows us to obtain magnetic
Reynolds number $Rm$ comparable to the ones used in the Maryland
experiment. In the presence of an external axial magnetic field, the system
exhibits most of the features obtained with a rotating outer sphere
and described in the previous section of this article. For instance,
when a magnetic field is applied to the hydrodynamical state, the
non-axisymmetric destabilizations of the equatorial jet can be
suppressed. Various non-axisymmetric states, different from this
equatorial jet instability, are also generated by the magnetic field.
In figure \ref{sis0}-right, we show the flow obtained when a
strong axial magnetic field is applied (with a
conducting inner sphere).

 In this case, the particular surface $\Sigma$ (which separates flow
 into two regions according to whether magnetic field lines are
 touching both spheres or only one of them) is located on the tangent
 cylinder: the fluid is at rest with the outer sphere outside the
 tangent cylinder, while the fluid inside the tangent cylinder rotates
 at $\Omega_i$ (for a conducting inner sphere) or at the intermediate
 rate $\Omega_i/2$ (for an insulating inner sphere). The spatial
 extension of the Ekman recirculation is also reduced. As noted
 before, this shear layer becomes unstable to non-axisymmetric
 perturbations for sufficiently large Reynolds number. For smaller
 values of the applied field, the return flow instability described in
 the previous section is also generated. \\

\begin{figure}

\centerline{
\hskip -4mm
\includegraphics[height=50mm]{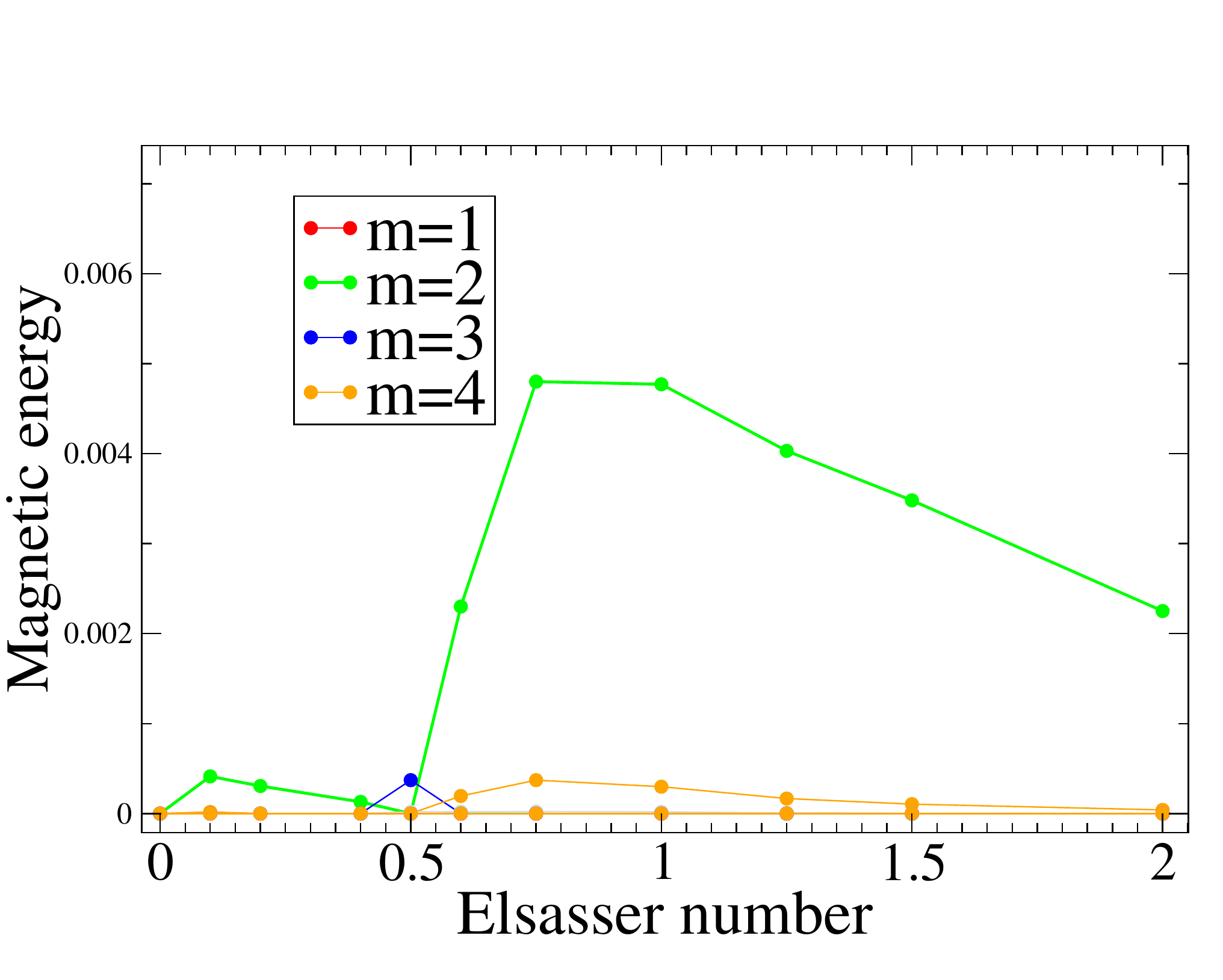}}
\vskip -1mm
\centerline{
\includegraphics[height=55mm]{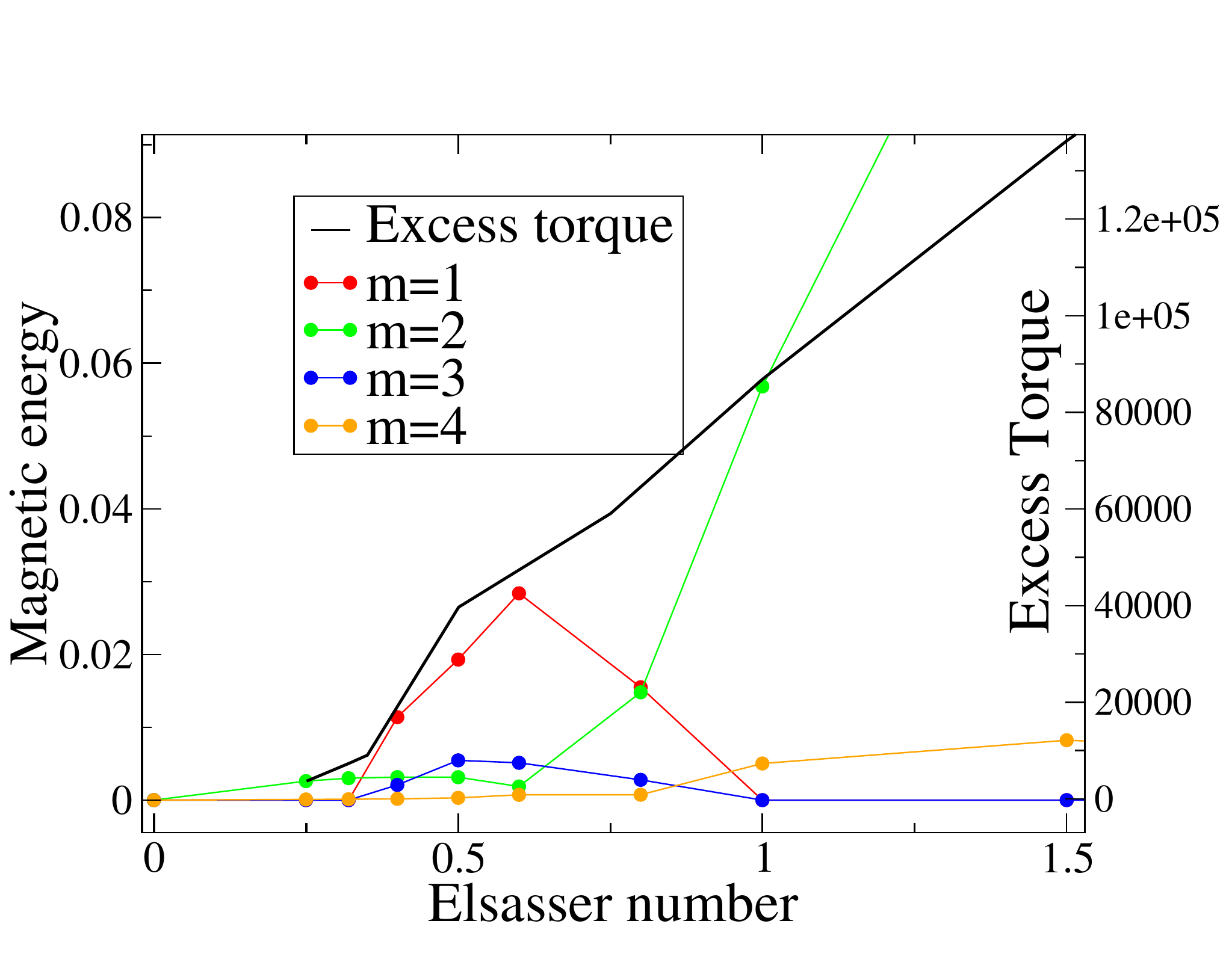}}

\caption{Bifurcation diagram of the magnetic energy when $\Lambda$ is
  increased, for $Re=5000$ and $Rm=50$, with an outer sphere at rest
  and an axial applied magnetic field (Maryland experiment
  configuration). Top: insulating inner sphere. Bottom: conducting
  inner sphere. In the latter case, a good agreement with the Maryland
  experiment is obtained, including the generation of an $m=1$ mode
  and increase of the torque on the inner sphere (to be compared with
  Fig.4 of \cite{Sisan06}).}
\label{siH}
\end{figure}

\begin{figure}
\centerline{
\includegraphics[height=45mm]{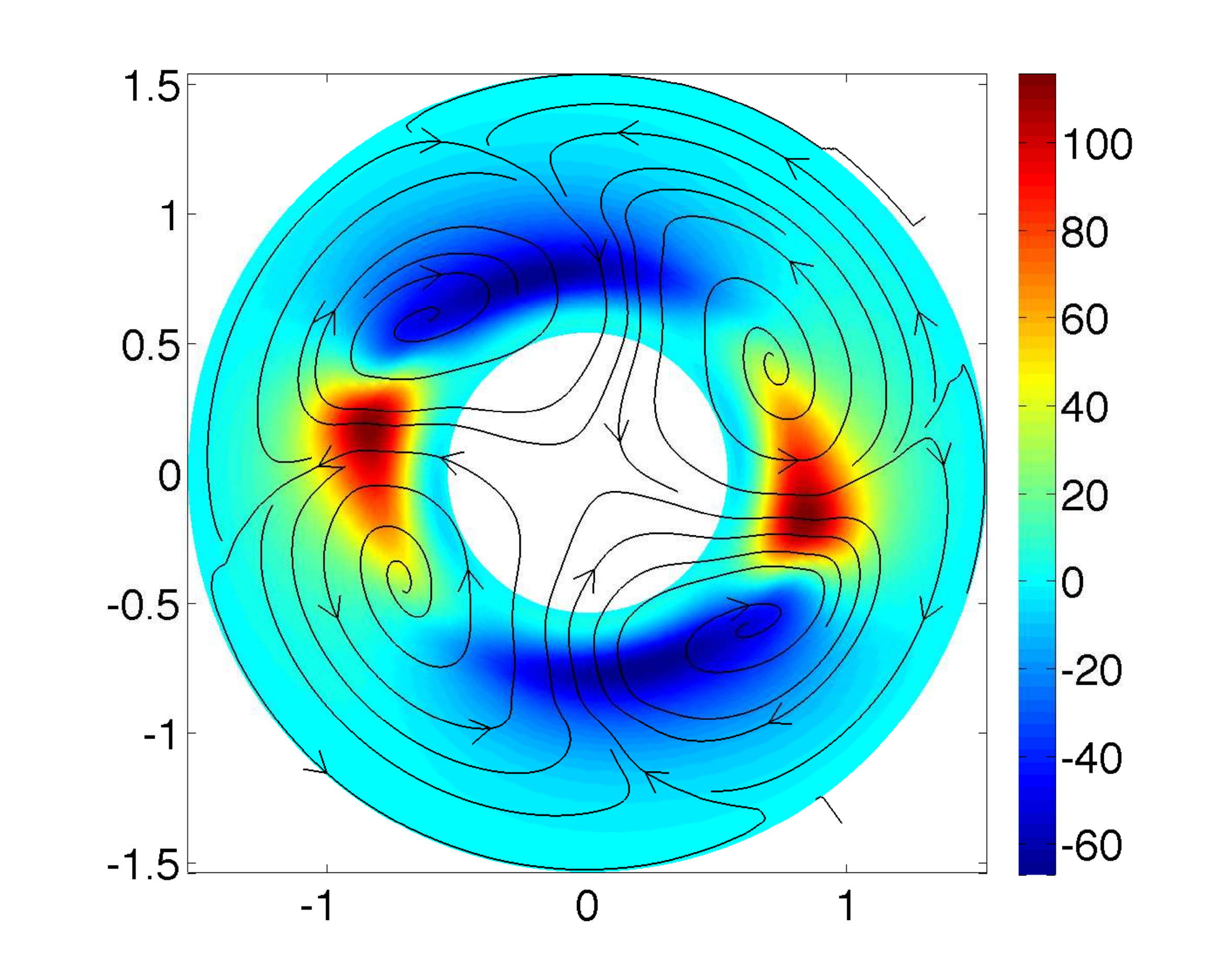} 
\includegraphics[height=45mm]{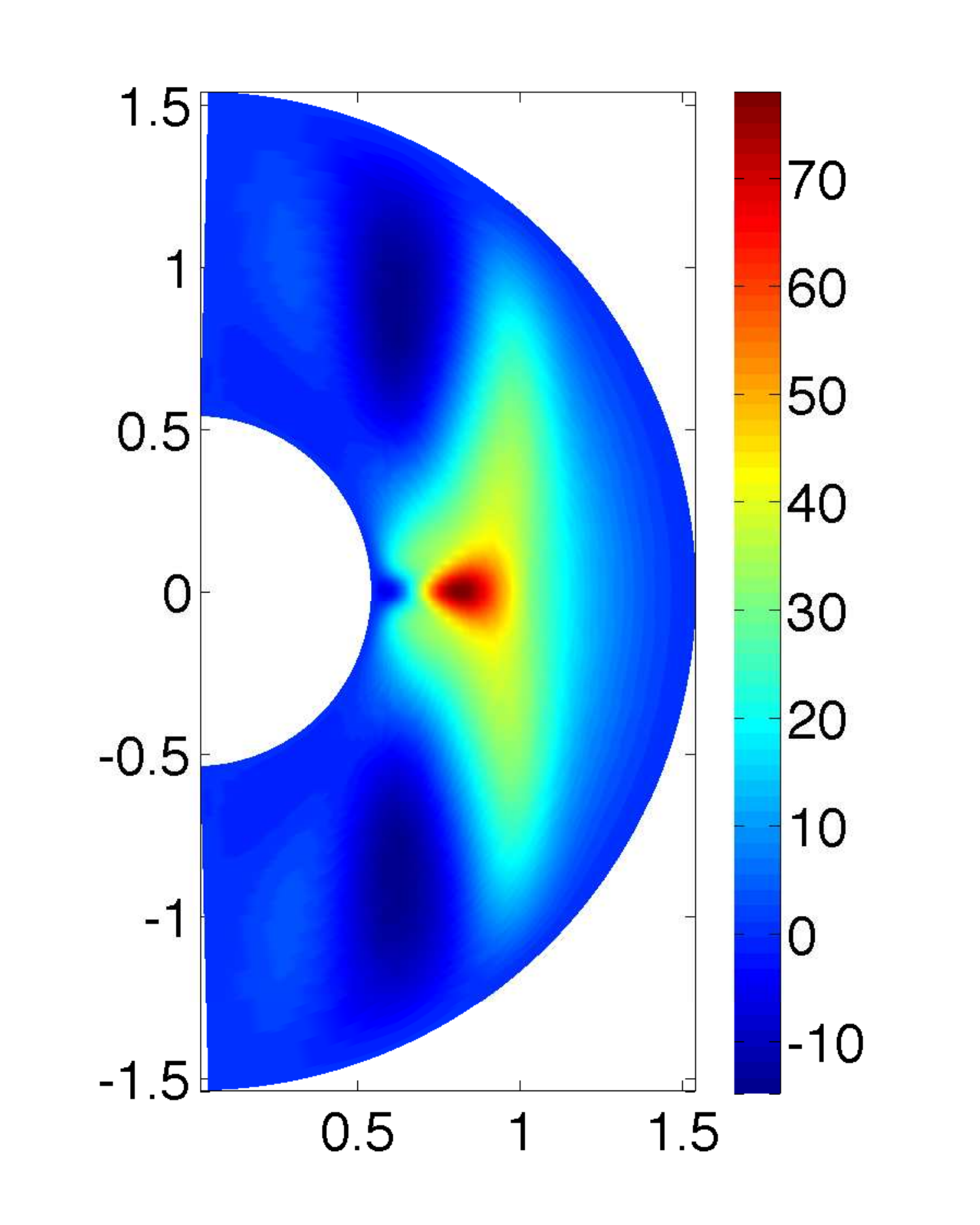}}
\centerline{
\includegraphics[height=45mm]{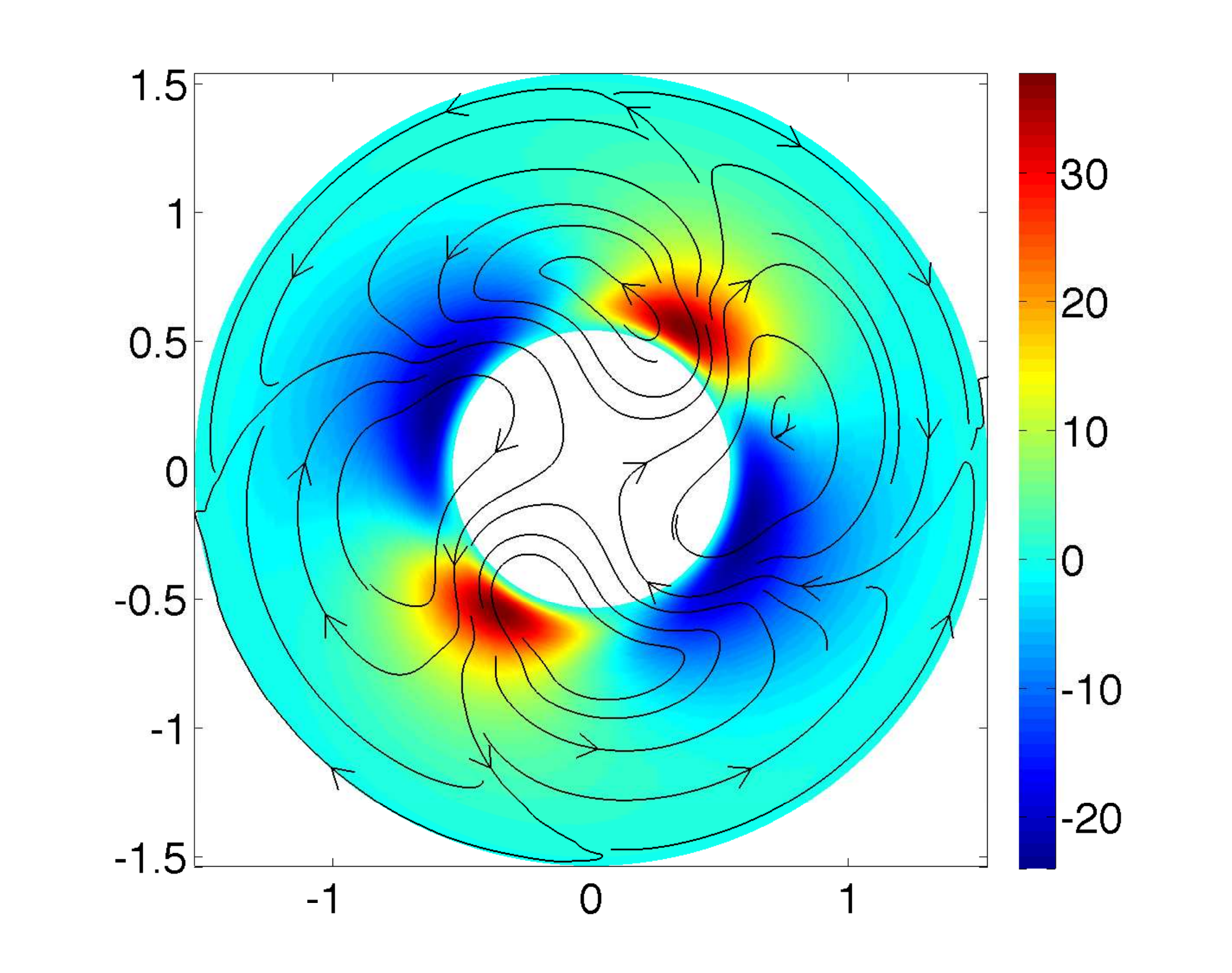} 
\includegraphics[height=45mm]{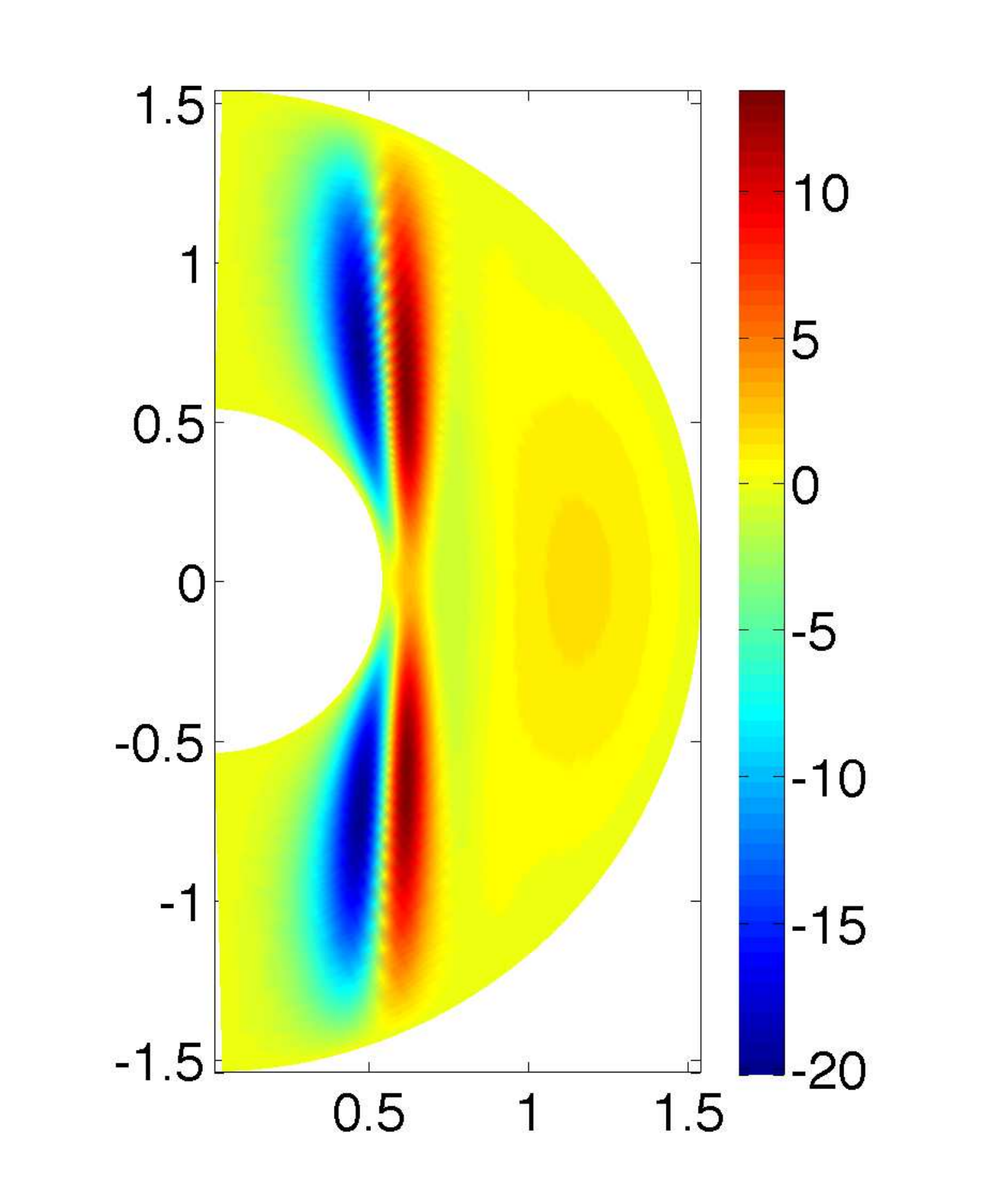}}
\caption{Structure of the MHD instabilities for $Re=5000$ with an
  insulating inner sphere and an axial field, formed of vortices in
  the horizontal plane. Figure shows the radial non-axisymmetric
  component of the velocity field $u_r$ in the equatorial plane (left)
  or in a given meridional plane at $\phi=0$(right). The two top
  snapshots show the $m=2$ mode obtained for $\Lambda=0.6$
  (instability related to the meridional return flow) and the two
  bottom snapshots show the field obtained for $\Lambda=2.$ (Shercliff
  layer instability).}
\label{shercliff_insu}
\end{figure}

 Figure \ref{siH} shows the evolution of the magnetic energy of
 different azimuthal modes for $Re=5000$ and $Rm=50$ as a function of
 the Elsasser number $\Lambda$. 

 In figure \ref{siH}-top, the inner sphere is insulating. In the whole
 range of Elsasser numbers explored here, the magnetic energy is
 dominated by an $m=2$ instability. At small Elsasser number
 ($\Lambda<0.5$), this corresponds to the hydrodynamical jet
 instability, equatorially antisymmetric, which extends into the
 magnetized regime. For $\Lambda>0.5$, a different instability occurs,
 which is symmetric with respect to the equator. Figure
 \ref{shercliff_insu} shows the structure of this instability for two
 different Elsasser numbers. As $\Lambda$ is increased, the
 oscillations gradually transits from a return flow instability
 associated with the meridional recirculation (top, $\Lambda=0.6$), to
 a Shercliff layer instability (bottom, $\Lambda=2$). In the latter
 case, the energy is concentrated on the tangent cylinder and consists
 of a series of vortices roughly independent of the $z$-direction.

\begin{figure}
\centerline{ \includegraphics[height=40mm]{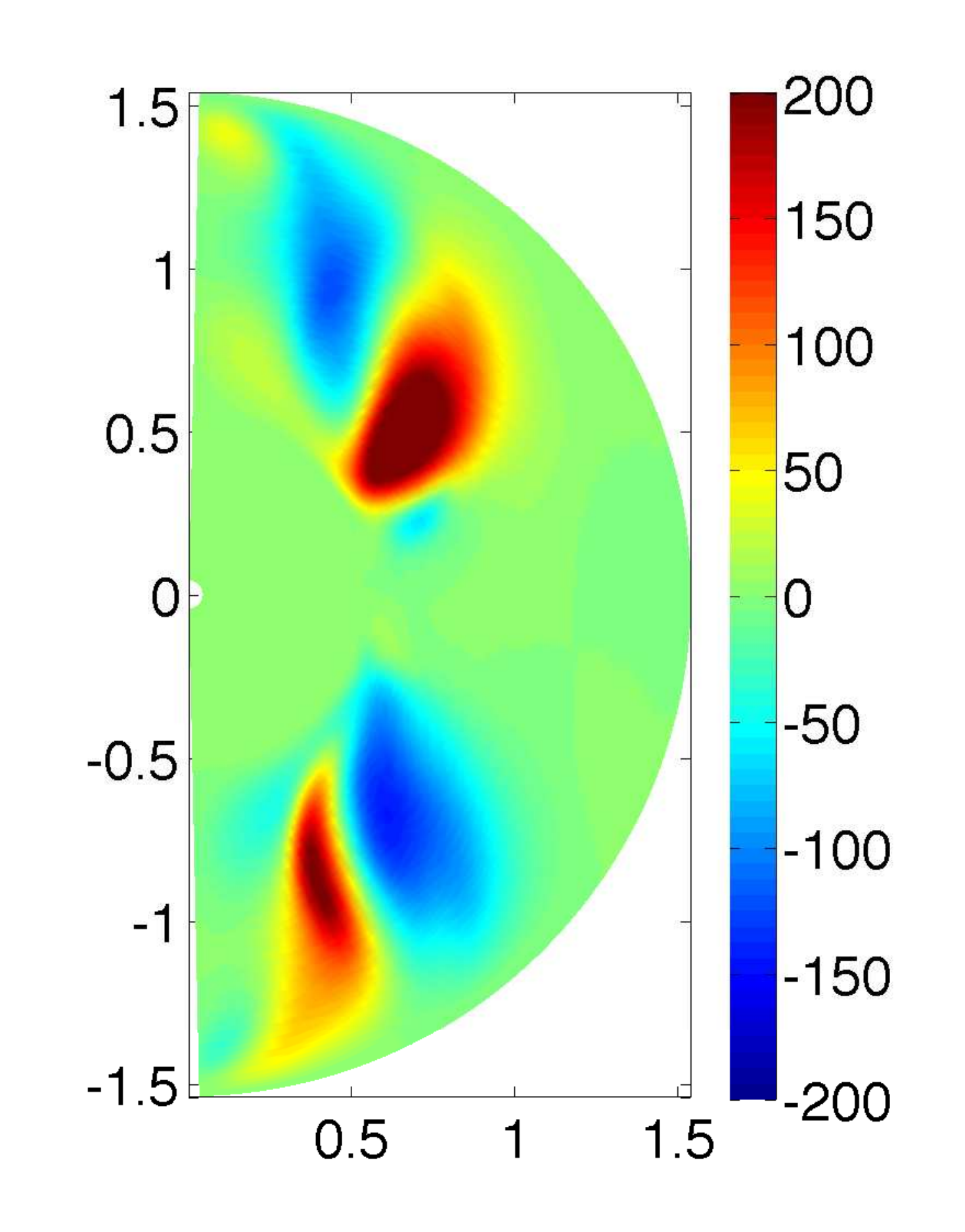}
             \includegraphics[height=40mm]{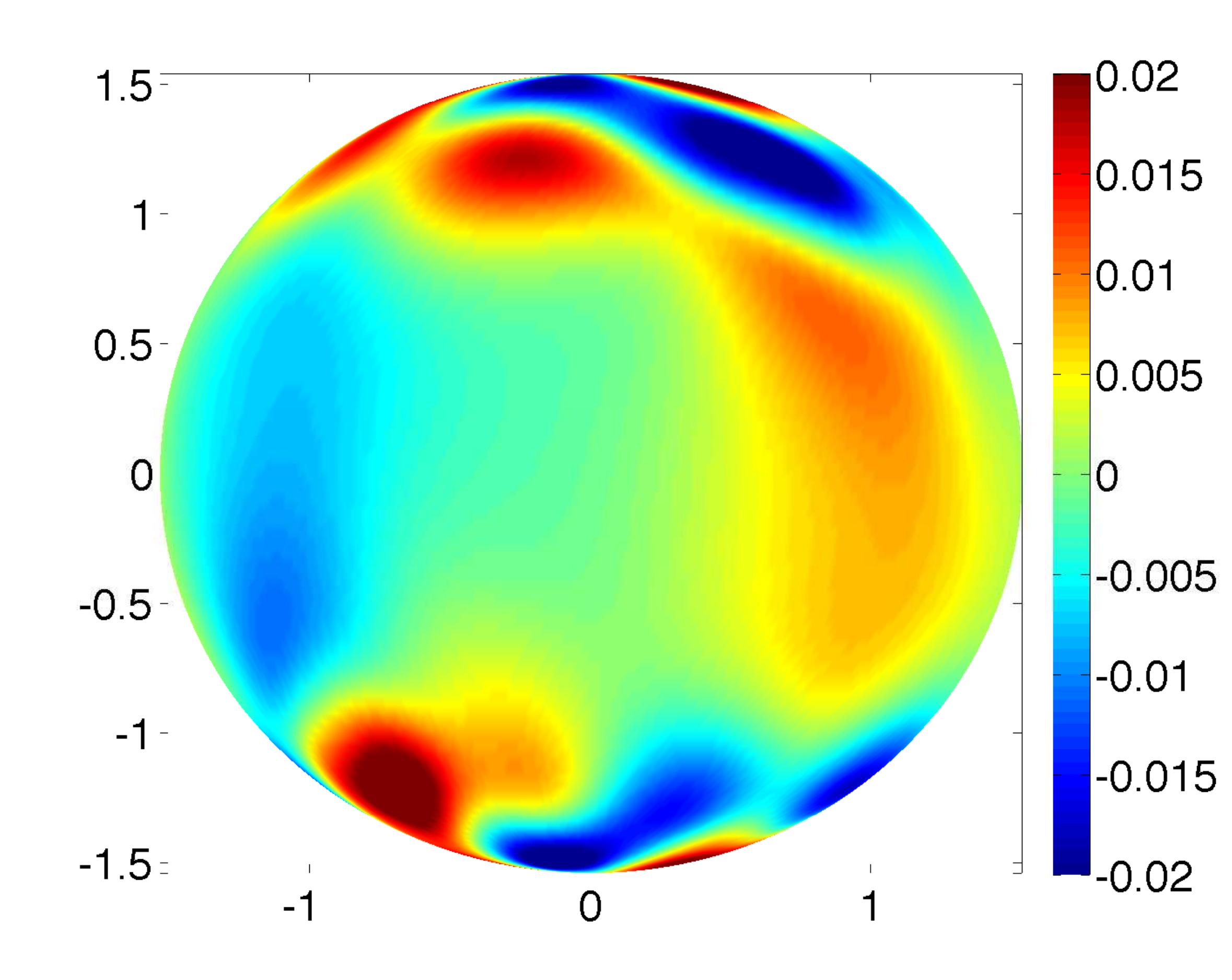}}
\caption{Structure of the $m=1$ instability obtained for
  $\Lambda=0.5$, $Re=5000$ and $Rm=50$, in the case of a conducting
  inner sphere. Left: non-axisymmetric $V_r$ in the
  meridional plane at $\phi=0$. Right: non-axisymmetric $B_r$ at the
  surface of the outer sphere. The structure is identical to the mode
  observed in the Maryland experiment, with the same equatorial and
  azimuthal symmetries.}
\label{m1Pm0p01}
\end{figure}

In figure \ref{siH}-bottom, the inner sphere is electrically
conducting. The electrical conductivity is identical to that of the
fluid, and thus smaller than the conductivity used in the Maryland
experiment, where an inner sphere made of copper is used. In spite of
this difference and of our lower Reynolds number, our numerical
simulations show a very good agreement with the results obtained in
the experiment (see for instance figure $4$ of \cite{Sisan06}): \\

- An $m=1$ mode is generated in a range of the Elsasser number similar
to the experiment. The generation of this $m=1$ mode
  is clearly due to the conductivity of the inner sphere. Note in
  particular that this instability is antisymmetric with respect to
  the equator, unlike Shercliff or return flow
  instabilities. However, the instability is still concentrated near
  the tangent cylinder(see figure \ref{m1Pm0p01}).  \\

- This non-axisymmetric mode can be suppressed by a strong field, and
a transition between different azimuthal wavenumbers is observed for
higher fields, in agreement with the Maryland experiment. \\

- The structure of the radial magnetic field, as shown in figure
\ref{m1Pm0p01}, possesses the same symmetry with respect to the
equator as the first mode obtained in the Maryland
experiment. At larger Elsasser number, this $m=1$
  instability is replaced by return flow instability and Shercliff
  layer instability, rather equatorially symmetric. This change of
  symmetry has also been reported in the Maryland experiment.

-The suppression of the $m=1$ mode and the generation
  of a smaller $m=2$ mode when the inner sphere is switched from
  conducting to insulating has also been observed in the Maryland
  experiment \cite{SisanThesis}. \\

- Finally, we computed the evolution of the excess torque applied to
the inner sphere in the conducting case. Like in the previous section,
as the flow becomes unstable to non-axisymmetric
perturbations, we observe a strong increase of the total torque
applied to the inner sphere (see the black curve in figure
\ref{siH}-bottom). This evolution has been interpreted as an
indication of MRI in the Maryland experiment. The fact that our
simulations reproduce this feature suggests that Shercliff layer
and return flow instabilities are also efficient
mechanisms to enhance the amount of angular momentum transported
outward. However, a large amount of the augmentation of our torque for
this conducting case and the one reported in the Maryland experiment
could be simply due to the attachment of magnetic field lines to the
inner conducting sphere. Indeed, analytical calculations of a rotating
conducting sphere with a uniform axial imposed field and surrounded by
an infinite medium of stationary Sodium, predict an important rise of
the torque \cite{Goodman}. Applied to our configuration, these
calculations lead to a torque of the same order of magnitude than the
one reported in the figure \ref{siH}. In any case, this shows that the
torque measured in the Maryland experiment cannot be assumed to be a
direct reflection of MRI instabilities.\\

\begin{figure}
\centerline{\includegraphics[height=40mm]{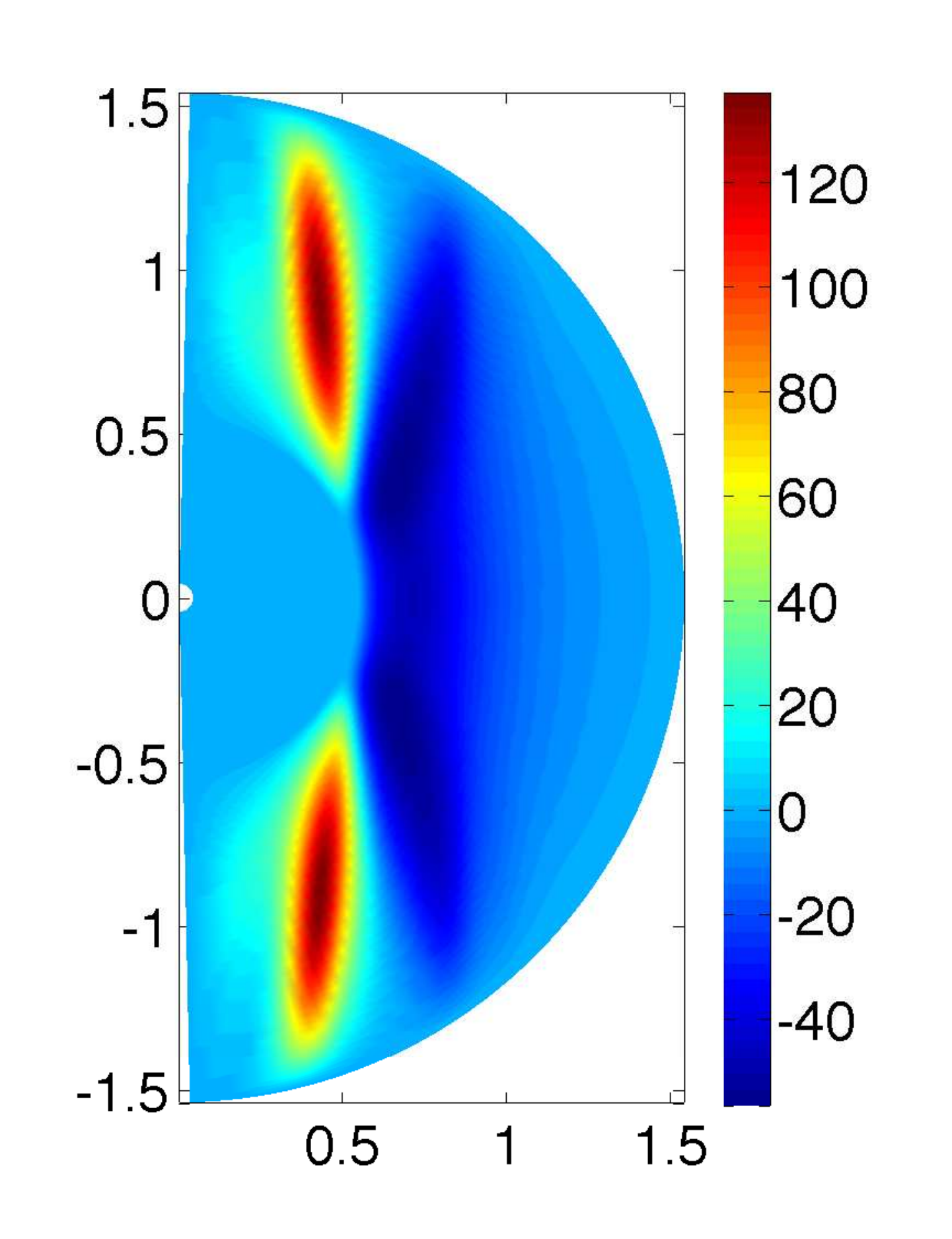}
            \includegraphics[height=40mm]{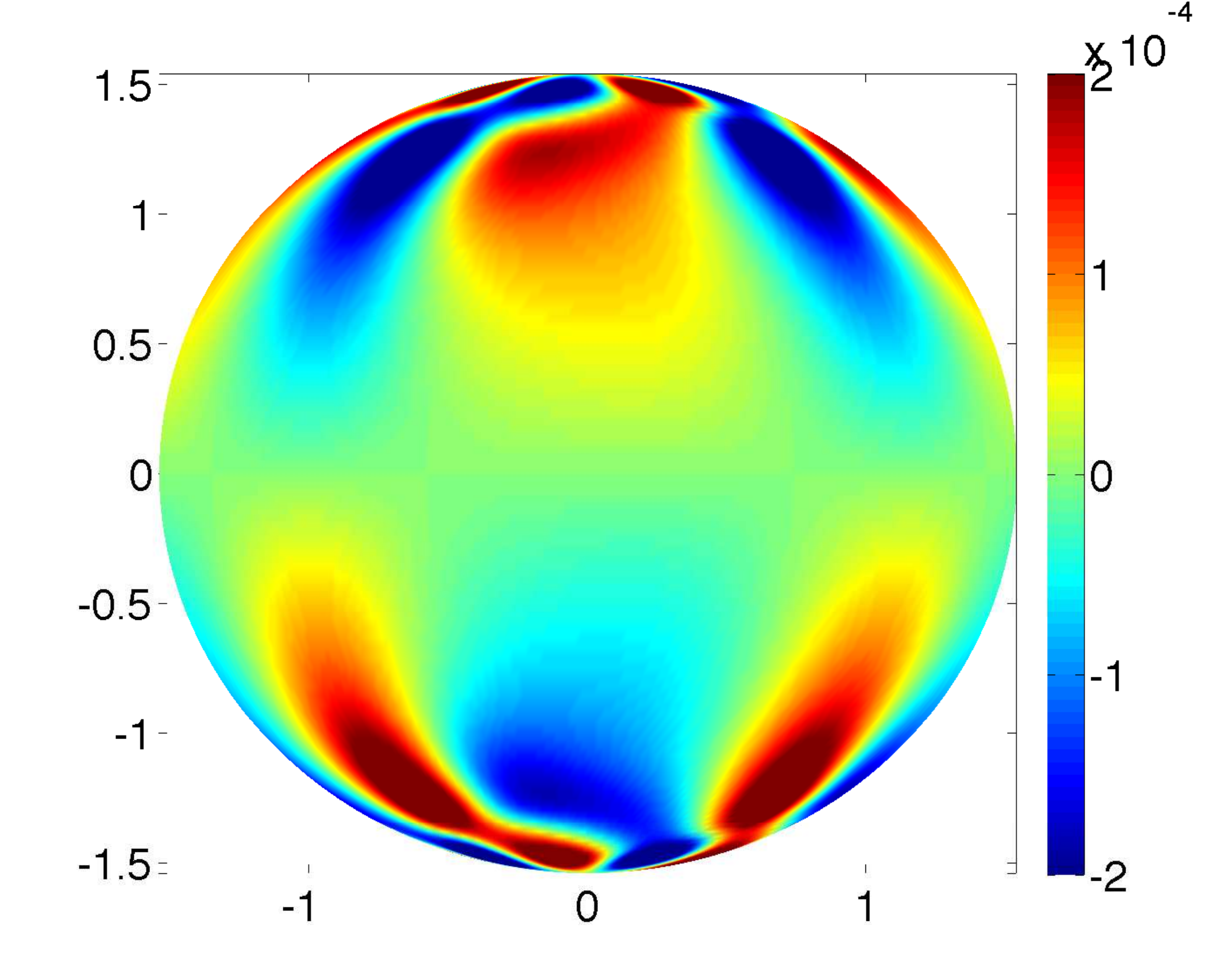}}
\caption{Structure of the non-axisymmetric component of the Shercliff
  layer instability for $Re=5000$, $Rm=0.5$ and $\Lambda=1.5$, with a
  conducting inner sphere. Left: non-axisymmetric $u_r$ in the
  meridional plane $\phi=0$. Right: non-axisymmetric $B_r$ at the
  surface of the outer sphere. Induction is not a necessary condition
  for the instability to be generated. }
\label{Rm0p5}
\end{figure}

An important observation is that both Shercliff
layer instability and return flow instability are
inductionless instabilities, in the sense that they can be generated
for arbitrary small magnetic Reynolds number if the hydrodynamical
Reynolds number is large enough.  Figure \ref{Rm0p5} shows for
instance the $m=2$ instability obtained for $Re=5000$, $Rm=0.5$,
$\Lambda=1.5$ and a conducting inner sphere. Note
  that this $m=2$ mode is in good agreement with inductionless
  calculations of \cite{Hollerbach09}, in which a similar setup is
  used (except for the magnetic boundary condition on the inner sphere
  which is insulating in \cite{Hollerbach09}).  This underlines the
difference of nature between the boundary driven instability reported
here and the standard MRI (for which induction is
necessary), despite the strong similarities between both
instabilities.

The several similarities between our simulations and the results of
the Maryland experiment emphasize the importance of considering the
role of boundaries in MRI experiment. These
numerical simulations indeed strongly suggest that the
non-axisymmetric modes observed in the experiment are destabilizations
of either the Shercliff layer (at large applied
  field) or the return flow (at smaller applied field), rather than
the MRI. These instabilities appear to be very robust, and extend in a
large region of the parameter space. This is therefore reasonable to
expect them to occur in the Maryland experiment, which uses a setup
similar to our numerical problem.

In addition to these non-axisymmetric modes, an $m=0$
  mode equatorially symmetric has also been reported in the Maryland
  experiment. It is interesting to mention that this axisymmetric
mode measured in the experiment has not been obtained in our numerical
simulation at $Pm<1$. However, for $Pm\sim 1$, axisymmetric modes are
generated in the simulations with an axial magnetic field, with or
without global rotation. They seem analogous to magnetostrophic MRI
modes, a modified version of the MRI in spherical geometry, possibly
relevant to the Earth core \cite{Petitdemange10} but relying on a
magnetostrophic equilibrium not achieved in our simulations. Their
analysis is beyond the scope of this paper, and will be studied in a
future work. This $m=0$ mode thus could be of a different nature than
the non-axisymmetric ones, and may not be related to the Shercliff
layer.

\section{Conclusion}

In this article, the flow of an electrically conducting fluid in a
spherical shell has been studied numerically. When a magnetic field is
applied to the system, two different effects are observed. First,
non-axisymmetric hydrodynamical bifurcations from the Stewartson layer
or from the equatorial jet can be suppressed by a sufficiently strong
magnetic field. But the applied field also has a destabilizing effect,
by either disrupting the axisymmetry of the
meridional return flow, or through a two-step process: first, an axial
or dipolar applied field creates a Shercliff layer, and second, this
MHD shear layer eventually becomes unstable to non-axisymmetric modes
if the magnetic field is not too strong.

We have seen that this Shercliff layer instability,
or the meridional return flow instability, are
dominated by different azimuthal wavenumbers depending on the
parameters $(Re,\Lambda)$ and the details of the configuration. These
instabilities are very important in the context of laboratory studies
of the magnetorotational instability. Indeed, they are relatively
robust, and share a lot of characteristics with the non-axisymmetric
MRI. The marginal stability curve is similar, with a destabilization
occurring only in a given range of the value of the applied
field. Whereas a finite value of the field is required to trigger the
instability, the free energy come from the azimuthal velocity and its
associated differential rotation (at least in the case of the
Shercliff layer instability). Moreover, we have shown that, like the
MRI, these instabilities yield an MHD turbulent state, and are very
efficient to transport the angular momentum outward. It is however
important to insist on the fact that Shercliff layer and return flow
instabilities are inductionless (unlike the standard
MRI), and rely on a different destabilization mechanism.

These similarities have important consequences for laboratory studies
of the MRI. First, as was suggested in
  \cite{Hollerbach09}, our numerical simulations strongly confirm
that results of the Maryland experiment are related to these boundary
driven instabilities rather than MRI.  A very good agreement is
obtained with experimental observations, including the sequence of
non-axisymmetric bifurcations, the geometry of the magnetic field and
the increase of the torque on the inner sphere. This work could also
have interesting echos for investigation of the MRI in a cylindrical
geometry. In this case, the finite geometry imposed by the
experimental approach makes it impossible to obtain an ideal Couette
flow or a quasi-Keplerian flow, because of the poloidal recirculation
created by the viscous stress at the vertical endcaps. This problem
has been circumvented by replacing the rigid endcaps at the top and
the bottom by two rings that are driven
independently. It has been shown that a flow profile
  in a short Taylor-Couette cell can be kept stable until $Re\sim
  10^{6}$ if the appropriate configuration of split end caps is used
  \cite{Ji06}. Similarly, the use of such split rings in the Promise
  experiment has led to a significant reduction of the Ekman pumping
  and a much clearer identification of the HMRI
  \cite{Stefani09}. Recent three-dimensional numerical simulations
\cite{Gissinger11} suggest that the jump of angular velocity between
inner and outer rings can be reinforced by applying an axial magnetic
field, and extended in the $z$ direction. This leads to the creation
of a Shercliff layer very similar to the one described in this paper,
which also undergoes some transition to non-axisymmetric modes. It is
thus possible that similar instabilities could be generated from the
Shercliff layer or from the poloidal return flow in these cylindrical
MRI experiments. This interpretation is reinforced by the fact that
the modes observed in the Princeton MRI experiment appear to be
inductionless \cite{Roach10}.

Finally, our numerical simulations show that investigation of the MRI
in the laboratory is significantly complicated by the presence of
no-slip boundaries. In spherical or cylindrical geometry, the applied
magnetic field interacts with these boundaries and can trigger MHD
instabilities very similar to the MRI. It could make very difficult
any distinction between these instabilities and MRI in an
experiment. The inductionless nature of Shercliff layer instability,
in contrast to the required induction for standard MRI, may be an
important key for the needed distinction between them.  However, as
pointed in the introduction, standard MRI continuously connects to the
Helical MRI (HMRI) \cite{Hollerbach95}, an inductionless version of
the MRI which can be regarded as an inertial oscillation destabilized
by resistive MHD in presence of an helical magnetic field
\cite{Kirilov10}. In this particular case, it would be interesting to
see how our boundary-driven instabilities can therefore be related to
global manifestations of the MRI.

\begin{acknowledgments}
This work was supported by the NSF under grant AST-0607472, by the
NASA under grant numbers ATP06-35 and APRA08-0066, by the DOE under
Contract No. DE-AC02-09CH11466, and by the NSF Center for Magnetic
Self-Organization under grant PHY-0821899.  We have benefited from
useful discussions with E. Edlund, A. Roach, E. Spence and
R. Hollerbach.
\end{acknowledgments}

%%%%%%%%%%%%%%%%%%%%%%%%%%%%%%%%%%%%%%
%%%%%%%%%%%% REFERENCES %%%%%%%%%%%%%%%%%%
%%%%%%%%%%%%%%%%%%%%%%%%%%%%%%%%%%%%%%

\end{document}